\documentclass[12pt,natbib]{aastex}
\usepackage{rotate}

\newcommand\4{{\footnotesize IV}}
\newcommand\3{{\footnotesize III}}
\newcommand\2{{\footnotesize II}}
\newcommand\1{{\footnotesize I}}
\newcommand\lam{{$\lambda$}}

\newcommand\kms{$\rm{km s^{-1}}$}

\newcommand\pd{\phantom{$-$}}
\newcommand\p{\phantom{:}}
\newcommand\pl{\phantom{$<$}}

\newcommand{\vrad}{$v_{\rm r}$}

\slugcomment{Accepted by ApJ}
\shorttitle{Blue supergiants in NGC\,3109}
\shortauthors{Evans et al.}

\begin{document}


\title{The ARAUCARIA Project: VLT-FORS spectroscopy of blue supergiants in NGC\,3109 -- 
Classifications, first abundances and kinematics\altaffilmark{1}}


\author{ C. J. Evans\altaffilmark{2}, F. Bresolin\altaffilmark{3}, M. A. Urbaneja\altaffilmark{3}, 
G. Pietrzy\'{n}ski\altaffilmark{4,5}, W. Gieren\altaffilmark{4} \\
and R.-P. Kudritzki\altaffilmark{3}
}

\altaffiltext{1}{Based on observations at the European Southern Observatory Very Large
Telescope in programme 171.D-0004.}
\altaffiltext{2}{UK Astronomy Technology Centre, 
                     Royal Observatory, 
                     Blackford Hill, 
                     Edinburgh, 
                     EH9 3HJ, UK}
\altaffiltext{3}{Institute for Astronomy, 
                     2680 Woodlawn Drive, 
                     Honolulu, HI 96822}
\altaffiltext{4}{Universidad de Concepci\'{o}n, 
                      Departamento de Fisica, 
                      Astronomy Group, 
                      Casilla 160-C, 
                      Concepci\'{o}n, 
                      Chile}
\altaffiltext{5}{Warsaw University Observatory, 
                     Al. Ujazdowskie 4, 
                     00-478, 
                     Warsaw, 
                     Poland}

\begin{abstract}
We have obtained multi-object spectroscopy of luminous blue
supergiants in NGC\,3109, a galaxy at the periphery of the Local Group
at $\sim$1.3\,Mpc.  We present a detailed catalog including
finding charts, $V$ and $I$ magnitudes, spectral classifications, and
stellar radial velocities.  The radial velocities are seen to trace
the rotation curves obtained from studies of the H~\1 gas.  From
quantitative analysis of eight B-type supergiants we find a mean
oxygen abundance of 12~$+$~log(O/H)~=~7.76 $\pm$0.07 (1-sigma
systematic uncertainty), with a median result of 7.8.  Given its distance, we
highlight NGC\,3109 as the ideal example of a low metallicity,
dark-matter dominated, dwarf galaxy for observations with the
next generation of ground-based extremely large telescopes (ELTs).

\end{abstract}


\section{Introduction}
\label{intro}

NGC\,3109 is a Magellanic-like spiral galaxy of type SBm \citep{dv} at
the outer edge of the Local Group, at a distance of 1.3~Mpc
\citep{cpb92,lee93,3109phot2}.  Its relative proximity and low foreground
reddening \citep{bh84} make it well-suited to exploring galaxy
formation and evolution in dwarf irregulars beyond the Magellanic
Clouds.  

Photometric studies have suggested that the main body of
NGC\,3109 has a low, SMC-like metal abundance \citep{ef85,g93}.
Moreover, from deep imaging, \citeauthor{g93} noted subtly
different populations across the disk of the galaxy that they
ascribed to different metallicities/star-formation histories in their
target fields, to differential reddening, or a combination of both.
The galaxy is thought to have a very metal-poor stellar halo
\citep{mza99}, although \citet{dbl03} have argued that the minor axis star 
counts can be explained by the inclination of the disk, rather than
requiring a significant halo population.  There also appears to be
evidence of interaction with the Antlia dwarf galaxy $\sim$1~Gyr ago
\citep{bdb01}.

Observations of the rotation curve of NGC\,3109 suggest 
a dominant dark-matter halo \citep{jc90}, that cosmological N-body
cold dark matter (CDM) simulations have struggled to reproduce
\citep{nfw}.  NGC\,3109 has also lent itself to tests of the
predictions of Modified Newtonian Dynamics \citep[MOND,][]{mil83}.
but observational uncertainties such as the adopted distance
led to differing conclusions \citep{lake89,mil91}.  Improved
observational data \citep[e.g.][]{bac01} have helped the situation, but
the debate continues \citep[e.g.][and references therein]{vrk06}.

As part of the larger ARAUCARIA project (P.I. Gieren) we have obtained
spectroscopy of some of the most visually-luminous stars in NGC\,3109,
using the Very Large Telescope (VLT) of the European Southern
Observatory (ESO).  These are the first spectroscopic observations of
resolved stars in this galaxy, enabling us to study the
kinematics and abundances of the young stellar population in this
dark-matter dominated dwarf.  The observations are detailed in
Section~\ref{obs}, followed by classification of the spectra in
Section~\ref{class}.  Stellar radial velocities are presented in
Section~\ref{rv}, in which we have compared them with H~\1 rotation curves, 
and in Section~\ref{abun} we undertake a quantitative analysis of
eight stars to investigate their chemical abundances.  Lastly, in 
Section~\ref{elt} we suggest NGC\,3109 as a target for future 
large telescopes.

\section{Target Selection and Observations}
\label{obs}

The data were taken with the FORS2 (focal-reducer and low dispersion
spectrograph) instrument at the VLT on 2004 February 24 \& 25.  The
multi-object spectroscopy with moveable slits (MOS) mode of operation
was used with the 600B grism for the blue-region observations and the
1200R grism in the red.  Targets were selected from $V$ and $I$-band
photometry obtained as part of the ongoing Cepheid variable search
programme at the Las Campanas Polish 1.3-m telescope \citep{3109phot}.
In conjunction with new $J$ and $K$-band VLT observations, these data
have been used to obtain a distance to NGC\,3109 of 1.30 $\pm$ 0.02
Mpc \citep{3109phot2}, the most precise distance estimate to date.

\begin{figure*}
\begin{center}
\includegraphics{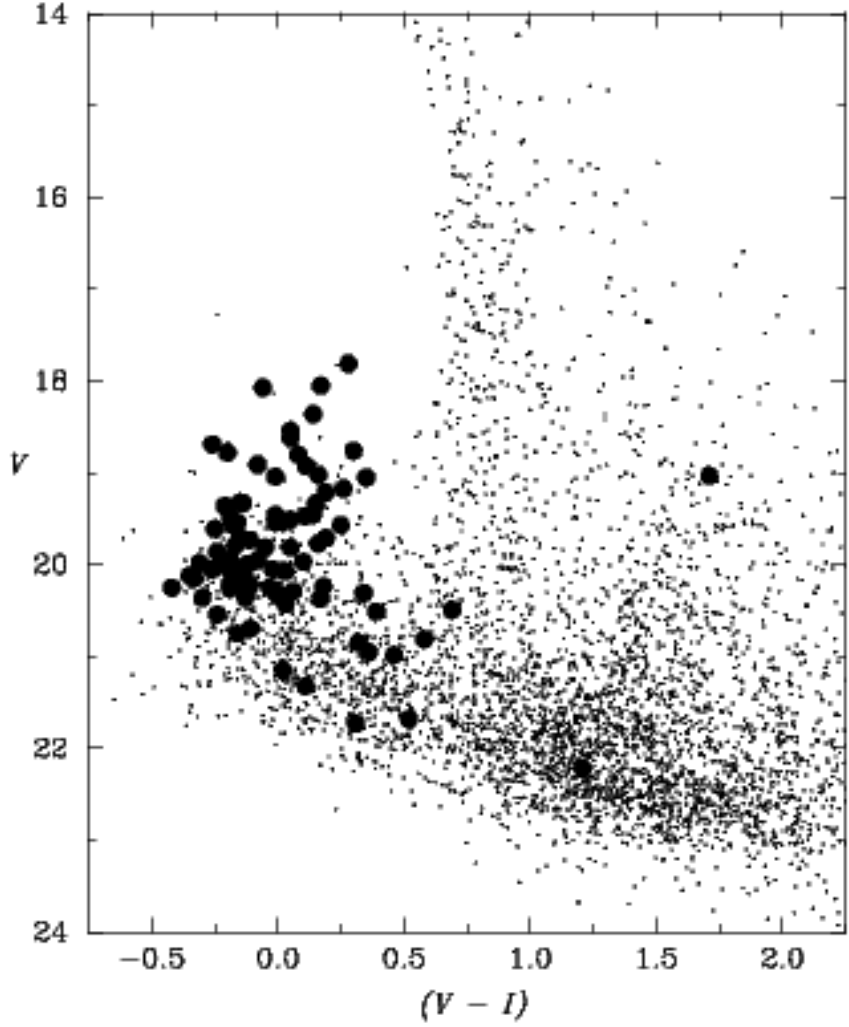}
\caption{Colour-magnitude diagram of VLT-FORS targets in NGC\,3109 (large closed circles), selected from  $V$ and $I$-band photometry \citep{3109phot}.}
\label{cmd}
\end{center}
\end{figure*}

Figure~\ref{cmd} shows the location of our targets in the colour-magnitude 
diagram.  Secondary targets were also observed serendipitously in
some slitlets; the spectra of these were also extracted when reducing
the data.  This is the origin of the redder targets in Figure~\ref{cmd}, in
particular the G-type supergiant with $(V-I) \sim$~1.7.

Finding charts from the FORS $V$-band pre-imaging are given in
Figures~\ref{targets1}, \ref{targets2}, and \ref{targets3}.  For
reference, the three strongly-saturated stars in Figure~\ref{targets1}
are those near the `F2' label in Figure~1 from \citet{b93}.  Moving
eastwards from the three reference stars, the FORS fields roughly span
the F2, F1, D2 and D1 fields from \citeauthor{b93}
Note the slight `drift' northeastwards in the
astrometry for the targets in the western part of Figure~\ref{targets2}. 
These were selected from a different FORS pre-image, but the offset of the 
target stars is evident from the isolated bright targets.  Stars \#4, 45 and 61 were
secondary targets that happened to be in the CCD gap in the pre-image.
Our fields also overlap with the imaging from \citet{g93}.  The western
quarter of Figure~\ref{targets1} is in their region C, and nearly all of 
our targets shown in Figure~\ref{targets2} are encompassed by their
region A.

Prior to extraction of the spectra, the {\sc l.a.cosmic} routines from
\citet{lacos} were used to clean the frames of cosmic rays.  The
spectra were then bias corrected, extracted and wavelength calibrated
using {\sc iraf}\footnote{{\sc iraf} is distributed by the National
Optical Astronomy Observatories, which are operated by the Association
of Universities for Research in Astronomy, Inc., under cooperative
agreement with the National Science Foundation.} (v2.12).  An optimal
extraction was used for the blue spectra, but we employed a simpler 
summed extraction routine for the red data -- we prefer this method
as nebular and/or intrinsic emission at H$\alpha$ can sometimes lead to
problems with sky-subtraction when using optimal extractions.  Indeed,
accurate sky subtraction is still challenging in those stars with
strong, nearby nebula emission.  Further manipulation of the
spectra was done with the Starlink package {\sc dipso}, including
correction to the heliocentric frame and rectification.  

The observations are summarised in Table~\ref{obsinfo}.  In
Table~\ref{targets} we list the observational details for each star,
by column these are: (1)~Running identifier; (2)~FORS field of
observation; (3)~Right Ascension; (4)~Declination; (5)~$V$ magnitude;
(6)~$V-I_{\rm c}$ colour; (7)~Spectral type; (8)~Radial velocity in
\kms; (9)~1-$\sigma$ uncertainty of radial velocity; (10)~Number of
absorption lines measured in determination of the radial velocity;
(11)~Signal-to-noise ratio of the spectrum near to H$\gamma$; 
(12)~Additional comments such as cross-identifications with other
catalogues \citep[e.g.][]{sc88}, or presence of strong nebular
emission lines.

The FWHM resolution (as defined by the arc lines) for the 600B data is
approximately 4.6~\AA, corresponding to 3.2 pixels.  For the 1200R
spectra this is 6.1~\AA\/ (3.1 pixels).  These both give a spectral
resolving power $R$ of $\sim$1,000.  As a consequence of the MOS
slitlets the exact wavelength coverage varies, but the \lam3650
to \lam5550~\AA\/ region is covered by all of the blue spectra.

\subsection{Photometric variability}\label{varsect}
The photometry from \citet{3109phot} was primarily obtained to monitor
Cepheid variables so we have 70 to 80 photometric observations of each
of our stars.  The data were acquired over two seasons, with
three-quarters of them from the first season (spanning Julian Dates
2452343 to 2452359).

The $V$ magnitudes and $(V-I_{\rm c})$ colours in Table~\ref{targets} are the
means of the multi-epoch values.  The expected scatter of the
photometric results at a given magnitude is illustrated by Figure~2
from \cite{3109phot}, in which their photometry is compared to that
from Hidalgo et al. (in preparation).  At $V \sim$~18 there is
excellent agreement, with an increasing scatter toward fainter
magnitudes (of order $\pm$0.1$^{\rm m}$ at $V =$~20). 

We find evidence for variability ($>$0.1$^{\rm m}$) in 5 of our targets
that are brighter than $V~=~$20, as shown in Figure~\ref{var}.  To give an indication of the
scatter of the individual measurements, the lightcurve of a
non-variable star (\#7) is also shown.  There may be small-scale
periodic (or aperiodic) variations in our other targets that are
undetected from qualitative inspection, but these are not explored
further here.

\section{Spectral Classification}
\label{class}
With previous studies suggesting that the metallicity of NGC\,3109 is
comparable to that of the SMC \citep[e.g.][]{ef85}, the FORS spectra were
classified by comparison to published SMC spectra.  For B-type
supergiants we use the SMC standards published by \citet{l97}, and
for A- and later-type stars we use the criteria from \citet{eh03} and
\citet{eh04}.  These sources have already tackled the issue of 
classification at low metallicities in the MK system.  For O-type
spectra we rely on the digital atlas of \citet{wf90}, with reference
to the SMC data presented by \citet{wal00}.  

In addition to the low foreground reddening, the total line-of-sight
reddening is also relatively low for stars in the major axis of
NGC\,3109, with E($B-V$) $\sim$0.14 \citep{d93}.  From the O star
spectra shown in Figure~\ref{ostars} 
it can be seen that there is some contamination of the spectra by
interstellar Ca~$K$ and $H$ (at \lam3933 and \lam3968 respectively, with 
the latter blended with H$\epsilon$ at \lam3970).  The intensity of the
interstellar components are comparable to those seen in some SMC spectra 
\citep[compare with the O-type spectra shown in Figure~2 from][]{eh04}, and
should not strongly affect the classifications for the A-type spectra -- particularly 
as the intensity of the metallic absorption lines is also considered when
classifying.

The resolution from FORS ($\sim$4.5~\AA\/ in the blue) is lower than
that usually used for precise classification.  We have taken this
into account when classifying the spectra (largely by degrading the
standards to the same effective resolution).  Classifications for our
targets are given in Table~\ref{targets}.

For spectra with signal-to-noise ratios of greater than 50, we show the
blue-region data (\lam3900 to \lam4750~\AA) in Figures~\ref{ostars} to
\ref{astars}.  To highlight the spectral sequence more clearly, there
is deliberate repetition of some spectra.

\subsection{Comments on individual stars}
We now give further explanation of our adopted spectral types for some
of the targets that display peculiar or interesting features, non-unique types etc.

\noindent{\it \#2 -- mid Be -- Figure~\ref{bmid}:}  
There is significant emission in the Balmer series in this spectrum.
The intensities of the He~\1 and Mg~\2 lines are consistent with a mid
B-type classification (of $\sim$B3-5), although of course there may be
infilling in the helium lines, and/or continuum effects from a
circumstellar disk.  There is weak [N~\2] emission in the wings of the
H$\alpha$ profile (see Figure~\ref{red1}), but the scale and extent of
the H$\alpha$ emission suggests a primarily non-nebular origin, hence the
Be-type classification.

\noindent{\it \#7 -- B0-1 Ia -- Figure~\ref{b0}:}
The uncertainty in this spectral type arises from what appears to be
very weak He~\2 \lam4542.  The spectrum is otherwise consistent with
that of a B1-type supergiant.

\noindent{\it \#9 -- B0.5 Ia -- Figure~\ref{b0}:} 
This bright star has a nearby companion (cf. Figure~\ref{targets1}) to
the extent that the spatial profiles were somewhat blended prior to
extraction.  The dominant spectrum appears to be that of a B0.5
supergiant, with weak He~\2 \lam4686 absorption.

\noindent{\it \#12 -- B8 Ia -- Figure~\ref{bmid}:}
The blue spectrum of star \#12 shows increasing amounts of infilling
in Balmer lines as one moves redwards, presumably from a stellar
outflow.  The H$\alpha$ profile displays broad and strong emission (with an
equivalent width of 7.5 $\pm$0.3~\AA).

\noindent{\it \#22 -- B1 Ia -- Figures~\ref{b0} \& \ref{bearly}:} 
From qualitative inspection the N~\2 \lam3995 line in this spectrum appears 
relatively strong for its type (but see comment in Section~\ref{abun}).

\noindent{\it \#24 -- B2-3 Iab -- Figure~\ref{bnon}:} 
The absence of Si~\3 combined with Mg~\2 absorption (albeit weak)
requires a type of B3.  However there is also weak absorption from O~\2
\lam4650 that would not usually be expected, suggesting a slightly earlier
type.

\noindent{\it \#35 -- O8-9.5 I(f) -- Figure~\ref{ostars}:} 
The strength of the He~\2 \lam4542 absorption is consistent with an
earlier type than the \lam4200 line, leading to a range of possible
classifications for this spectrum.  Weak N~\3 emission is seen at
\lam\lam4634-40-42, with infilled He~\2 \lam4686.

\noindent{\it \#36 -- mid B -- Figure~\ref{bmid}:} 
P Cygni emission is seen in both the H$\gamma$ and H$\beta$ lines;
similarly at H$\alpha$ (Figure~\ref{red1}).  The ratio of Si~\2 
\lam\lam4128-32 to He~\1 \lam4121 requires a mid to late B-type classification 
for this spectrum, well matched by the other He~\1 lines.  However
the weakness of the Mg~\2 \lam4481 feature contradicts this somewhat,
suggesting a slightly earlier type.

\noindent{\it \#59 -- B0-3 -- Figure~\ref{bnon}:}
Weak absorption is seen from Si~\3 but a precise
spectral type cannot be assigned due to the absence of other diagnostic
lines (such as Mg~\2 \lam4481, the CNO blend at $\sim$\lam4650 etc).

\noindent{\it \#91 -- B1-2e (shell) -- Figure~\ref{bnon}:}
The Balmer lines in this spectrum are very narrow, resembling those
seen in shell stars.  The presence the \lam4650 blend suggests an
early B-type, but we lack other diagnostic lines to assign a unique
type (e.g. Si~\3, Si~\4, Mg~\2).  With moderate signal-to-noise
observations of low metallicity targets, this is a common problem in
this spectral domain \citep[c.f. star \#59 above, and][]{eh04}.  This
star appears to be heavily reddened, although it is at the faint end
of the photometric survey and so a relatively large degree of
uncertainty in the colour can probably be expected \citep[cf.][]{3109phot}.

\section{Stellar Radial Velocities}
\label{rv}
Radial velocities, $v_{\rm r}$, were calculated from the means of
manual measurements of the line centres of absorption lines.  Where
possible the primary Balmer lines (H$\beta$, H$\gamma$, H$\delta$, H8
and H9) were supplemented by measurements of He~\1 and He~\2 lines,
and occasionally from metallic lines (e.g. in the luminous B-type
supergiants).  We do not use the H$\epsilon$ line owing
to its blend with the Ca~{\it H} line.

The mean $v_{\rm r}$ values for each of our targets
are given in Table~\ref{targets}, together with
1-$\sigma$ uncertainties and the number of lines used
(dependent on both the spectral type of the star, and the data
quality).  To investigate the spatial variation of the radial
velocities we initially binned our results.  The median 1-$\sigma$
(internal) uncertainty is 19~\kms, but we adopt
a more conservative bin-size of 50~\kms (thereby also providing
a reasonable number of stars per bin).
In Figure~\ref{rvfig} we show the spatial distribution of the stars in
each of four bins, i.e. \vrad $<$ 350, 350 $<$ \vrad $<$
400, 400 $<$ \vrad $<$ 450, and \vrad $>$ 450~\kms.  Although
limited by both sample size and velocity precision, Figure~\ref{rvfig}
shows a  trend of increasing velocities with increasing right ascension, 
i.e. rotation of the galaxy as traced by the young stellar population.

In Figure~\ref{rvc} we plot differential velocities, $\Delta v$
(i.e. $v_{\rm r} - v_{\rm sys}$), as a function of radius along the
main disk of the galaxy, taking the systemic velocity ($v_{\rm sys}$)
as 402~\kms\/ \citep{bac01}.  We have corrected the
velocities in the figure for an inclination of 75$^{\rm o}$ \citep{jc90}.  The
radius is calculated by finding the radial distance of each star from
the optical centre of the galaxy
\citep[$\alpha =$~10$^{\rm h}$03$^{\rm m}$06\fs6, $\delta
=$~$-$26$^{\rm o}$09$'$32$''$, J2000.0;][]{bac01}, projected onto the
major axis of the galaxy.  The position angle of NGC\,3109 is 93$^{\rm
o}$ \citep{jc90}, so the projection term related to this component is relatively
minor.  In Figure~\ref{rvc} we also plot rotation curves from H~\1
observations (\citeauthor{jc90}, solid line, shifted by 2~\kms~so that the
systemic velocities tally) and H$\alpha$ (\citeauthor{bac01}, dotted line).  

In general, the stars track the H~\1 and H$\alpha$ rotation curves, but
with a fair degree of scatter.  Further observations would be
of significant value to determine whether the stellar results are
revealing genuine sub-structures in the disk, or if we are simply
limited by the small sample (and/or undetected nebular contamination at
our relatively low spectral resolution).

\section{Stellar abundances in NGC\,3109}\label{abun}

Early B-type supergiants have numerous metallic absorption lines that
are useful for abundance determinations.  Here we analyse eight
supergiants from our sample using the {\sc fastwind} model atmosphere code \citep{fw1,fw2}.
The methods for quantitative analysis at the spectral resolution from FORS are
described in detail elsewhere \citep{maup1,maup2}.
In brief, each model is described by an effective temperature (T$_{\rm eff}$), a surface gravity and a
stellar radius, each of which is defined at $\tau_{\rm Ross} =$~2/3.
Additional model parameters are the exponent of the wind velocity law
($\beta$), the microturbulent velocity ($v_{\rm turb}$), the mass-loss
rate ($\dot{M}$), the wind terminal velocity ($v_\infty$) and a set
of chemical abundances.  Terminal velocities are taken from the
spectral type--$v_{\infty}$ relation from \citet{kp00}.

To limit the number of free parameters, we follow the approach described
by \citet{maup33}, namely fitting the effective temperature, surface gravity, global
metallicty and $Q'$ (which, at fixed $v_\infty$, takes into account the mass-loss rate 
and stellar radius).  In calculation of the atmospheric structure
$v_{\rm turb}$ is fixed to 10~\kms, but this is then varied in the spectral synthesis to
obtain uniform abundances from different lines of the same metallic species.

The extinction and absolute magnitudes of our targets 
are given in Table~\ref{ext}.  The colour excess, $E(V-I_{\rm c})$, 
was calculated from the observed colours
(Table~\ref{targets}) and intrinsic colours obtained from the {\sc
fastwind} models, convolved with the appropriate filter.  Some
iteration is required here, but once the stellar radius is
approximately correct, the colours do not vary significantly.  The
$V$-band extinction (A$_{V}$) was found using the relations
from \citet{ccm89} to calculate the (Cousins) $I$-band extinction.
Absolute magnitudes (M$_V$) were then calculated taking
the distance modulus to NGC\,3109 as 25.57 \citep{3109phot2}.  Note that
non-physical (i.e. negative) colour excesses were obtained for two of
our targets (\#7 and \#22), likely due to nearby H~\2 regions.  For
these two stars the mean extinction from \citet{3109phot2} of
$E(B-V) =$~0.09 was adopted.

Effective temperatures, microturbulent velocities, and silicon
abundances were obtained from fits to the Si~\2/\3/\4 lines, with
gravities and wind mass-loss rates obtained from fits to the Balmer
lines, principally H$\beta$.  The H$\alpha$ line is a better
diagnostic of the stellar wind, but at the distance of NGC\,3109 the
nebular emission varies on very small scales and sky-subtraction
becomes critical.  Even with the FORS slitlets, there is evidence for
over/under-subtraction in many of the H$\alpha$ profiles (cf. the
nebular lines), and we do not use H$\alpha$ in our analysis.  The 
associated uncertainties of this method are T$_{\rm eff} \pm~$1,000~K 
and log$g \pm$~0.1 dex \citep[see discussion by][]{maup33}.

The primary abundance diagnostics for nitrogen and oxygen
are: N~\2 \lam\lam3995, 5050, 5100; O~\2 \lam\lam4072-76, 4317-19,
4414-16.  There are also many weaker lines and the
optimum model is determined by attempting to reproduce as many
features as possible -- the results are included in
Table~\ref{results}.  The dependence of the chemical abundances on the
adopted stellar parameters is shown for two of our stars in
Tables~\ref{abunerr22} and \ref{abunerr37}.  The combined
uncertainties from different parameters (added in quadrature) are
shown in the final columns, with individual abundances robust to $\pm$0.2~dex.

We also estimate abundances for Mg~\2 \lam4481 and Si~\3 \lam\lam4553-72.
These are less secure than the nitrogen and oxygen abundances due to
the smaller number of lines and their relative weakness.  Moreover, we do not quote
carbon abundances for our stars.  The primary isolated line of carbon is 
C~\2 \lam4267, which is barely visible in the FORS spectra, notwithstanding the
fact that there are many problems regarding the use of this line in theoretical models 
\citep[e.g.][]{l03}.  The rich blend of CNO lines around \lam4650 can provide an 
indirect tracer of the carbon abundance, but from data at this spectral resolution
the combined uncertainties render any value as very unreliable.

A summary of the physical parameters for the B-type supergiants is
given in Table~\ref{results}.  The final {\sc fastwind} models are
compared with the FORS spectra in Figures~\ref{fits1} and
\ref{fits2}, in which the model spectra have been smoothed to
the effective resolution of the observations.  In Figures~\ref{err_22}
and \ref{err_37} we show models for stars \#22 and \#37 in which the
metallic abundances are $\pm$0.2~dex compared to the adopted values -- in
some cases individual O~\2 lines are better fit by the increased or
decreased values, but we consider the fit to multiple lines in our
determination of the final abundance.

The mean oxygen abundance is 12~$+$~log(O/H)~=~7.76 $\pm$0.07 (1-sigma
systematic uncertainty).  This is in excellent agreement with results
from H~\2 regions.  \citet{lee03a} give an oxygen abundance of 7.73
for one target, and \citet{lee03b} found a mean of 7.63 for a further
five regions.  Moreover, new results from $\sim$10 compact H~\2
regions find a mean abundance of 7.74 $\pm$ 0.10 (Dr.~M. Pe\~{n}a,
private communication).  The mean oxygen abundance found here is also
comparable to recent results for stars in the WLM galaxy \citep{wlm}.
We note that oxygen abundances are typically considered in 0.1~dex
increments between models, so the mean values here give the impression
of unrealistic precision.  The median value for our 8 stars is 7.8
(cf. 7.9 from the three stars in WLM).  The results are significantly
lower than those found for stars in the SMC, for which
log(O/H)$+$12~=~8.13 and 8.14 for B- and A-type supergiants
respectively \citep{tl04,tl05,venn99}.

Our abundance estimates for magnesium and silicon are more in keeping
with those found in the SMC stars from \citet{tl05}.  However, in many
cases our estimates are upper limits, and the exact abundance of the
alpha-elements will require higher-resolution spectroscopy.  

Where of note, we now provide brief comments on specific stars:

\noindent{\it \#11:}
The O~\2 feature at \lam\lam4072-76 is somewhat distorted, with some
uncertainty in definition of the continuum.  Only upper limits are
possible for both nitrogen and magnesium as there are not significant
lines from either element.  The mass-loss rate was fixed to reproduce
H$\beta$, with T$_{\rm eff}$ from the ionization balance of He~\1/\2
and Si~\3/\4.

\noindent{\it \#7:}    
The spectrum of \#7 suffers from significant emission from a nearby H~\2 
region.  Both H$\alpha$ and H$\beta$ are strongly contaminated, with H$\gamma$
and H$\delta$ also in-filled somewhat.  The gravity for this star was obtained from the
higher-order Balmer lines, that appear well reproduced.  T$_{\rm eff}$ was found
from the balance of He~\1/\2 and Si~\3/\4.  Again the nitrogen abundance is an
upper limit.

\noindent{\it \#9:}
The mass-loss rate was fixed to reproduce H$\beta$, although the line-core is 
not well matched -- H$\alpha$ shows some nebular contamination, it is likely that
H$\beta$ is also infilled slightly.

\noindent{\it \#22:}
The strong absorption close to \lam3995 is not consistent with being
from N~\2 when the other nitrogen features in this spectrum are
considered -- the observed absorption is also very slightly
red-shifted and its origins remain unclear.

\noindent{\it \#37:}
The ionization balance of Si~\2/\3/\4 is used to find T$_{\rm eff}$ --
note that this star is a `transitional' type in which neither the
Si~\2 nor the Si~\4 lines are strong.

\section{NGC\,3109 as a target for future extremely large telescopes}
\label{elt}

Plans for the next generation of large ground-based telescopes (the
so-called extremely large telescopes, or ELTs) are now gaining
momentum.  Excluding the big spiral galaxies, the total stellar mass in
NGC\,3109 is larger than in other Local Group systems -- in the
context of ELTs, NGC\,3109 presents an exciting opportunity to study
many stages of stellar evolution in a very metal-poor environment.
With a 30-m primary aperture, good signal-to-noise spectra could be
obtained down to approximately M$_{V} = -$2.  Such observations would
enable detailed studies of the young, massive population (i.e. O- and early
B-type stars on the main-sequence) and of stars on the asymptotic giant
branch (AGB).  The fainter stars could be observed at spectral
resolutions comparable to those from FORS, with higher resolutions
($R~\sim$~20,000) employed to investigate wind parameters and
abundances in, for example, main-sequence O-type stars.

Moreover, with less demanding signal-to-noise requirements, an ELT
would be able to trace the kinematics of the non-supergiant population
(via e.g. the calcium triplet).  Crucial input to cosmological
simulations could be obtained from observations along the major and
minor axes of NGC\,3109, to fully investigate the structure of this
dark-matter dominated dwarf.

We also note that most of the current plans for ELT instruments
include some degree of adaptive optics (AO) correction
\citep[e.g.][]{crc06}.  For observations in NGC\,3109, a `seeing-limited' 
optical instrument should be adequate, e.g. the Wide Field Optical
Spectrograph (WFOS) proposed for the Thirty Meter Telescope (TMT)
project \citep[e.g.][]{pfm06}.  However, at larger distances and finer
spatial scales, the AO systems will most likely restrict observers to near-IR
wavelengths.  Quantitative studies of OB-type stars in this domain
have advanced in recent years \citep[e.g.][]{len04,rp05,hkk05}.  To
fully exploit the likely capabilities of an ELT, continued efforts are
required in this region -- both in terms of atomic data
\citep[e.g.][]{p05}, and comparison studies in the Milky Way and
Magellanic Clouds.

\section{Summary}

We have presented spectra from an exploratory survey of the young,
massive-star content of NGC\,3109.  These are the first spectral
observations of the young stellar population in this galaxy.  Although
of limited resolving power, the spectra have been used to estimate
stellar radial velocities, which appear largely consistent with
published rotation curves from observations of the gas.  Our stellar
oxygen abundances agree with recent results from H~\2 regions, and
are significantly lower than those found for stars in the SMC 
\citep[e.g.][]{tl05}.  They are also comparable to those in the 
WLM galaxy \citep{wlm}.  We require higher-resolution spectroscopy to 
obtain precise abundances for the alpha-elements, but it is clear
that the stars in NGC 3109 have metal abundances that are very 
deficient when compared to the solar neighbourhood, and likely
even lower than in the SMC.

\section{Acknowledgements}
CJE acknowledges financial support from the UK Particle Physics and
Astronomy Research Council (PPARC).  GP and WG gratefully acknowledge
financial support for this work from the Chilean Center for Astrophysics, under
grant FONDAP 15010003.  We thank the referee for their constructive comments.

\bibliographystyle{aa}
\bibliography{ms}

\clearpage

\begin{center}
\begin{deluxetable}{cccc}
\tabletypesize{\small}
\tablewidth{0pc}
\tablecolumns{4}
\tablecaption{Summary of observations \label{obsinfo}}
\tablehead{
\colhead{Field} & Grism & \colhead{Date} & \colhead{Exposure time}\\
&&&[sec]
}
\startdata
1 & 600B & 2004-02-24 & 4$\times$2700 \\
2 & 600B & 2004-02-25 & 4$\times$2400 \\
3 & 600B & 2004-02-24 & 3$\times$2700$+$1$\times$2400 \\
4 & 600B & 2004-02-25 & 1$\times$2700$+$2$\times$2400 \\
1 & 1200R & 2004-02-24 & 4$\times$2500 \\
2 & 1200R & 2004-02-25 & 4$\times$2400 \\
\enddata
\end{deluxetable}
\end{center}

\clearpage

\begin{center}
\begin{deluxetable}{llccccllllll}
\tabletypesize{\scriptsize}
\tablewidth{0pc}
\tablecolumns{12}
\tablecaption{Observational parameters of target stars \label{targets}}
\tablehead{
\colhead{Star} & \colhead{Field} &
\colhead{$\alpha$} & \colhead{$\delta$} & \colhead{$V$} & \colhead{$V-I_{\rm c}$} & 
\colhead{Spectral Type} & \colhead{$v_{\rm r}$} & \colhead{$\sigma$} & \colhead{\# lines} & \colhead{S/N} & 
\colhead{Comment} \\
&&(J2000.0)&(J2000.0)&&&&(\kms)&(\kms)&&&}
\startdata
 1 &  2 & 10 03 11.93 & $-$26 09 29.01 & 17.81 &   \pd0.28 & A2 Ia & 412 & 9 & 5 & 195 &  \\
 2 &  1 & 10 03 27.47 & $-$26 10 06.41 & 18.05 &   \pd0.17 & mid Be  & 477 & 9 & 6 & 180 & SC88-B155\\
 3 &  1 & 10 03 17.66 & $-$26 10 00.67 & 18.07  & $-$0.06   & B1 Ia  & 419 & 24 & 9 & 110 &  SC88-B80\\
 4 &  2 & 10 03 16.74 & $-$26 09 22.90 & 18.36 &    \pd0.14 & B9 Ia & 454 & 25 & 8 & 100 & SC88-B75\\
 5 &  2 & 10 03 05.37 & $-$26 08 56.58 & 18.54 &    \pd0.05 & B8 Ia & 411 & 23 & 10 & 95 & SC88-B56\\
 6 &  2 & 10 02 52.99 & $-$26 09 51.65 & 18.62 &    \pd0.05 & B8 Ia & 370 & 8 & 10 & 140 &  SC88-B34\\
 7 &  2 & 10 02 54.69 & $-$26 08 59.64 & 18.69 &   $-$0.26 & B0-1 Ia &  382 & 19 & 10 & 110 & SC88-B138\\
 8 &  3 & 10 02 53.15 & $-$26 09 37.03 & 18.76 &    \pd0.30 & A7 II & 393 & 19 & 5 & 95 & SC88-B36 \\
 9 &  2 & 10 02 49.77 & $-$26 08 45.04 & 18.78 &   $-$0.20 & B0.5 Ia & 375 & 16 & 9 & 140 & SC88-B136\\
 10 & 1 & 10 03 26.63 & $-$26 08 54.24 & 18.80 &    \pd0.08 & B: $+$em/neb? & 464 & 20 & 5  & 65 & SC88-B120\\
 11 & 1 & 10 02 58.61 & $-$26 09 50.95 & 18.91 &   $-$0.08 & B0 I &  354 & 22 & 7 & 125 & SC88-B148\\
 12 & 1 & 10 03 07.09 & $-$26 09 41.26  & 18.92 & \pd0.11 & B8 Ia & 399 & 15 & 7  & 80 & SC88-B57\\
 13 & 2 & 10 03 08.51 & $-$26 09 57.31 & 19.02 &    \pd0.16 & A0 Iab & 405 & 12  & 5 & 130 & SC88-B67\\
 14 & 2 & 10 02 59.54 & $-$26 09 12.44 & 19.03 &    \pd1.71 & G2 I & 317  & 18 & 3 & $-$ & SC88-R14\\
 15 & 2 & 10 03 13.10 & $-$26 09 35.61 & 19.04 &   $-$0.01 & B5 Ia & 419  & 17 & 12  & 120 & SC88-B90\\
 16 & 2 & 10 03 10.06 & $-$26 09 48.53 & 19.05 &    \pd0.35 & A7 II & 381 & 17  & 5 & 60 & \\
 17 & 1 & 10 03 20.21 & $-$26 06 44.62 & 19.17 &    \pd0.26 & A3 II  & 447 & 23 & 5  & 50 &  DKI1874\\
 18 & 2 & 10 02 57.76 & $-$26 08 33.25 & 19.21 &    \pd0.19 & B1 I & 392 & 29 & 6 & 105 & SC88-B104\\
 19 & 3 & 10 02 49.67 & $-$26 09 21.95 & 19.32 &   $-$0.15 & B1 Ia  & 363 & 17 & 11 & 85 &SC88-B33\\
 20 & 2 & 10 03 03.22 & $-$26 09 21.41 & 19.33 &   $-$0.14 & O8 I & 407  & 11 & 8 & 70 & \\
 21 & 3 & 10 03 09.96 & $-$26 08 27.11 & 19.33 &    \pd0.15 & A0 Iab & 397 & 20 & 5 & 50 & \\
 22 & 2 & 10 02 59.88 & $-$26 09 12.60 & 19.36 & $-$0.21 &  B1 Ia  & 391 & 25 & 14 & 65 & \\
 23 & 1 & 10 03 19.95 & $-$26 09 55.01 & 19.44 &    \pd0.14 & A1 Ib  & 453 & 18 & 5 & 130 & SC88-V14a \\
 24 & 1 & 10 03 01.89 & $-$26 09 01.04 & 19.46 &   $-$0.01 & B2-3 Iab & 346 & 15 & 11 & 80 & \\
 25 & 1 & 10 03 22.95 & $-$26 10 30.91 & 19.47 &    \pd0.11 & A0 Ib  & 429 & 7 & 5 &  90 & \\
 26 & 2 & 10 02 46.32 & $-$26 09 46.88 & 19.48 &   $-$0.19 & B3 Ib & 337 & 13 & 11 & 140 & \\
 27 & 2 & 10 03 01.85 & $-$26 09 08.91 & 19.49 &    \pd0.00 & B2.5 Ia  & 420 & 17 & 10 & 90 &  \\
 28 & 1 & 10 03 08.51 & $-$26 09 21.56 & 19.51 &    \pd0.05 & B2.5 Iab  & 390 & 30 & 11 & 90 &  \\
 29 & 1 & 10 03 12.50 & $-$26 10 14.81 & 19.54 &    \pd0.02 & B8 Ia  & 421 & 13 & 8 &  75 & \\
 30 & 2 & 10 02 55.53 & $-$26 09 54.82 & 19.54 &   $-$0.01 & A0 Iab & 400 & 21 & 7 & 85 & \\
 31 & 3 & 10 02 55.55 & $-$26 10 03.57 & 19.54 &   $-$0.16 & B1-2 Ib & 409 & 34 & 10 & 60 & \\
 32 & 1 & 10 03 03.41 & $-$26 08 46.06 & 19.57 &    \pd0.25 & A3 II & 370 & 8 & 5 & 95 & SC88-V7,65a\\
 33 & 3 & 10 03 02.45 & $-$26 09 36.11 & 19.57 &   $-$0.18 & O9 If & 334 & 33 & 5 & 60 & \\
 34 & 1 & 10 03 14.24 & $-$26 09 16.96 & 19.61 &   $-$0.25 & O8 I(f)  & 409 & 25 & 10 & 90 & \\
 35 & 3 & 10 03 13.65 & $-$26 09 55.76 & 19.70 &   $-$0.15 & O8-9.5 I(f) & 476 & 16 & 7 & 60 & \\
 36 & 1 & 10 03 29.56 & $-$26 08 39.96 & 19.71 &    \pd0.19 & mid B  & $-$ & $-$ & $-$ & 80 & P Cyg em. \\
 37 & 2 & 10 02 47.33 & $-$26 09 45.07 & 19.73 &   $-$0.11 & B2 Iab  & 364 & 22 & 13 & 90 & \\
 38 & 2 & 10 02 56.57 & $-$26 09 55.58 & 19.77 &    \pd0.16 & B1-3 III & 379 & 30 &8&50&\\
 39 & 4 & 10 02 46.97 & $-$26 10 14.34 & 19.79 &   $-$0.17 & O9.5 II: & 337 & 33 & 6 & 70 &  \\
 40 & 2 & 10 03 14.72 & $-$26 09 57.40 & 19.81 &   $-$0.05 & B8 Ib  & 430 & 13 & 5 & 90 & \\
 41 & 3 & 10 03 11.85 & $-$26 10 08.64 & 19.81 &    \pd0.05 & A0 Iab & 454 & 28 & 4 & 55 & \\
 42 & 3 & 10 03 01.36 & $-$26 08 26.69 & 19.86 &   $-$0.24 & B0-2 & 428 & 22 & 5 & 55 & \\
 43 & 4 & 10 03 11.73 & $-$26 10 18.00 & 19.94 &   $-$0.19 & O9.5 I  &  424 & 24 & 6 & 40 & \\
 44 & 3 & 10 02 58.59 & $-$26 09 58.04 & 19.95 &   $-$0.08 & early-B Ib & 394 & 33 & 7 & 35 & \\
 45 & 4 & 10 03 16.55 & $-$26 09 38.46 & 19.97 &    \pd0.10 & B8-A0 I & 441 & 20 & 5 & 25 & \\
 46 & 4 & 10 02 55.61 & $-$26 08 58.35 & 19.98 &   $-$0.31 & early-B & 394 & 15 & 9 & 40 &  \\
 47 & 3 & 10 02 55.09 & $-$26 10 07.25 & 19.99 &   $-$0.12 & B2.5 Ib  & 408 & 28 & 6 & 65 & \\
 48 & 4 & 10 02 56.20 & $-$26 08 58.23 & 19.99 &   $-$0.19 & late-O If  & $-$ & $-$ & $-$ & 40 & str. H{\tiny II} \\
 49 & 3 & 10 02 59.69 & $-$26 09 24.38 & 20.00 &   $-$0.22 & O9 II & 390 & 29 & 5 & 40 &\\
 50 & 1 & 10 03 10.62 & $-$26 09 22.56 & 20.05 &   $-$0.02 & B8: Ib & 399 & 10 & 3 & 80 &  \\
 51 & 4 & 10 02 54.45 & $-$26 09 03.69 & 20.05 &   $-$0.26 & early-B (B0?) & 351 & 20 & 5 & 60 & \\
 52 & 3 & 10 02 51.78 & $-$26 09 42.14 & 20.07 &    \pd0.03 & A0 Ib & 381 & 22 & 5 & 45 & \\
 53 & 4 & 10 02 51.40 & $-$26 09 35.02 & 20.07 &   $-$0.16 & B0: Ib & 405  & 35 & 5 & 55 &   \\
 54 & 3 & 10 02 46.61 & $-$26 09 12.45 & 20.10 &   $-$0.18 & B1-3 II-Ib & 352 & 23 & 5 & 45 &\\
 55 & 4 & 10 03 04.04 & $-$26 09 10.30 & 20.12 &   $-$0.35 & B0-2 & 413  & 19 & 6 & 45 & str. H{\tiny II} at H$\beta$\\
 56 & 4 & 10 02 46.16 & $-$26 09 55.26 & 20.14 &   $-$0.17 & early-B & $-$ & $-$ & $-$ & 65 &   \\
 57 & 4 & 10 02 47.12 & $-$26 10 14.27 & 20.15 &   $-$0.33 & early-B &  372 & 22 & 6 & 45 &  \\
 58 & 4 & 10 02 53.27 & $-$26 09 48.92 & 20.17 &   $-$0.13 & B9: Iab  & 385  &10 & 5 & 35 &  \\
 59 & 1 & 10 03 15.31 & $-$26 09 24.64 & 20.21 &   $-$0.19 & B0-3 & 446 & 15 & 10 &  75 & \\
 60 & 3 & 10 03 08.17 & $-$26 10 18.11 & 20.22 &   $-$0.17 & B0-0.5 &  434 & 31 & 8 & 35 &  \\
 61 & 3 & 10 03 16.54 & $-$26 10 20.96 & 20.23 &   $-$0.11 & early-B? & $-$ & $-$ & $-$ & 60 & str. H{\tiny II} \\
 62 & 3 & 10 03 14.60 & $-$26 10 06.00 & 20.24 &    \pd0.18 & A5 II & 475 & 19 & 5 & 50 & SC88-V9,45,57a \\
 63 & 1 & 10 03 00.10 & $-$26 09 28.72 & 20.25 &   $-$0.42 & late-O & $-$ & $-$ & $-$ & 55 & str. H{\tiny II} \\
 64 & 2 & 10 02 51.43 & $-$26 08 47.06 & 20.25 &   $-$0.03 & B0-2 & 565 & 23 & 8 & 55 & \\
 65 & 4 & 10 03 13.39 & $-$26 08 03.31 & 20.26 &   $-$0.19 & early-B & 490 & 16 & 4 & 35 & shell star? \\
 66 & 4 & 10 03 01.39 & $-$26 08 01.48 & 20.29 &    \pd0.06 & late-O & $-$ & $-$ & $-$ & 40 & str. H{\tiny II} \\
 67 & 1 & 10 03 24.64 & $-$26 09 35.03 & 20.31 &    \pd0.34 & A5 II & 454 & 11 & 5 & 50 &   \\
 68 & 4 & 10 03 02.57 & $-$26 09 30.70 & 20.33 &    \pd0.00 & A0 Iab:  & 412  & 15 & 5 & 30 & \\
 69 & 4 & 10 03 07.76 & $-$26 09 17.48 & 20.36 &   $-$0.13 & early-mid B & 408 & 33 & 6 & 30 &  \\
 70 & 4 & 10 02 58.97 & $-$26 08 11.58 & 20.36 &   $-$0.30 & B5 Ia & 424 & 33 & 5 & 40 & SC88-V6,25,31a\\
 71 & 4 & 10 03 14.15 & $-$26 10 06.71 & 20.37 &    \pd0.17 & B9-A0 Ib: & 405 & 30 & 4 & 35 &  \\
 72 & 4 & 10 02 49.73 & $-$26 09 42.95 & 20.40 &    \pd0.03 &  A0 Ib:  & 394 & 29 & 5 & 40 &   \\
 73 & 4 & 10 02 57.30 & $-$26 09 50.10 & 20.43 &    \pd0.03 & A0: Iab  &  392 & 11 & 5 & 35 &  \\
 74 & 3 & 10 03 05.77 & $-$26 09 11.36 & 20.44 &     $-$ & B0-2 II: & 440  & 17 &  5 & 40 & \\
 75 & 2 & 10 03 14.34 & $-$26 09 57.82 & 20.49 &    \pd0.69 & F2 I & 504 & 12 & 4 & $-$ & \\
 76 & 4 & 10 03 09.41 & $-$26 09 49.43 & 20.51 &    \pd0.39 & B9-A0 Ib  & 370 & 23 & 4 & 35 & \\
 77 & 3 & 10 03 05.03 & $-$26 09 23.78 & 20.52 &     $-$ & late-O & $-$ & $-$ & $-$ & 40 & str. H{\tiny II} \\
 78 & 2 & 10 02 49.47 & $-$26 08 45.18 & 20.54 &   $-$0.24 & O9.5 III  & 402 & 25 & 9 & 70 & \\
 79 & 2 & 10 03 05.83 & $-$26 09 06.92 & 20.55 &     $-$ & B9 Ib & 407 & 11 & 4 & 40 &  \\
 80 & 1 & 10 03 12.06 & $-$26 10 14.97 & 20.70 &   $-$0.11 & early-B (B1-2)  & 398 & 18 & 8  & 40 & \\
 81 & 2 & 10 02 47.00 & $-$26 09 45.21 & 20.75 &   $-$0.16 & B1: Ib  & 389 & 19 & 10 & 65 & \\
 82 & 3 & 10 03 10.90 & $-$26 10 08.34 & 20.81 &    \pd0.58 & F2: I & 382  & 13 & 3 & $-$ &  \\
 83 & 2 & 10 03 05.00 & $-$26 08 56.13 & 20.85 &    \pd0.32 & A3 II  & 296 & 17 & 5 & 45 & \\
 84 & 2 & 10 03 02.43 & $-$26 09 21.49 & 20.95 &    \pd0.36 & A5 II & 393 & 27 & 5 & 30 & \\
 85 & 4 & 10 03 05.67 & $-$26 09 21.20 & 20.98 &    \pd0.46 & early-B &  408 & 33 & 4 & 25 &  \\
 86 & 2 & 10 02 46.72 & $-$26 09 47.27 & 21.15 &    \pd0.02 & late-B II & 445 & 12 & 3  & 35 &  \\
 87 & 3 & 10 03 07.81 & $-$26 10 17.77 & 21.32 &    \pd0.11 & B & 292 & 30 & 3 & 20 & SC88-V36c \\
 88 & 1 & 10 03 05.60 & $-$26 09 17.03 & 21.41 &     $-$ & early-mid B & 455 & 14 & 5 & 30 & \\
 89 & 1 & 10 03 00.81 & $-$26 09 28.20 & 21.68 &    \pd0.52 & A5: II & 263 & 22 & 4 & 20 & \\
 90 & 2 & 10 02 59.31 & $-$26 09 12.72 & 21.73 &    \pd0.31 & A7 II: &  482 & 14  & 3 & 25 & \\
 91 & 1 & 10 03 27.18 & $-$26 10 06.11 & 22.22 &    \pd1.21 & B1-2e (shell)  & 440 & 18 & 9 &  55 & \\
\enddata
\tablerefs{DKI \citep{dki85}; SC88 \citep{sc88}}
\end{deluxetable}
\end{center}

\begin{center}
\begin{deluxetable}{llcccccc}
\tabletypesize{\footnotesize}
\tablewidth{0pc}
\tablecolumns{8}
\tablecaption{Extinction and absolute magnitudes for target stars \label{ext}}
\tablehead{
\colhead{ID} & \colhead{Sp. Type} & \colhead{$V$} & \colhead{$V-I_{\rm c}$} & \colhead{E$(V-I_{\rm c})$} & 
\colhead{A$_{V}$} & \colhead{$E(B-V)$} & \colhead{M$_{V}$} }
\startdata
11 &  B0 I    &   18.91 & $-$0.08 &  0.16      &  0.41 &  0.13 &   $-$7.07 \\ 
7  &  B0-1 Ia &   18.69 & $-$0.26 & $-$0.04\pd &  0.27 & 0.09 &   $-$7.15 \\
9  &  B0.5 Ia &   18.78 & $-$0.20 &  0.01      &  0.03 & 0.00 &   $-$6.82 \\
3  &  B1 Ia   &   18.07 & $-$0.06 &  0.10      &  0.26 & 0.08 &  $-$7.76  \\
22 &  B1 Ia   &   19.36 & $-$0.21 & $-$0.02\pd &  0.27 & 0.09 &   $-$6.48 \\
37 &  B2 Iab  &   19.73 & $-$0.11 &  0.04      &  0.10 & 0.03  &   $-$5.94 \\
27 &  B2.5 Ia &   19.49 & \pd0.00  &  0.15      &  0.39 & 0.13  &   $-$6.47 \\
28 &  B2.5 Iab&   19.51 & \pd0.05 &  0.20      &  0.52 & 0.17  &   $-$6.58 \\
\enddata
\tablecomments{The observed colour excesses for stars \#7 and \#22 are non-physical, 
due to likely contamination by nearby H~\2 regions.  The adopted $E(B-V)$ is the mean 
from \citet{3109phot2} }
\end{deluxetable}
\end{center}

\clearpage
\begin{center}
\begin{deluxetable}{llcccccccccccc}
\tabletypesize{\scriptsize}
\tablewidth{0pc}
\tablecolumns{14}
\tablecaption{Physical parameters of target stars \label{results}}
\tablehead{
\colhead{ID} & \colhead{Sp. Type} & \colhead{T$_{\rm eff}$} & \colhead{log$g$}  &
\colhead{R} & \colhead{M$_{\rm spec}$} & \colhead{log(L/L$_{\odot}$)} & 
\colhead{$v_{\rm turb}$} & \colhead{$Y_{\rm He}$} & 
\colhead{$\epsilon_{\rm N}$} & \colhead{$\epsilon_{\rm O}$} &
\colhead{$\epsilon_{\rm Mg}$} & \colhead{$\epsilon_{\rm Si}$} & \colhead{[O/H]} 
\\
& & [kK] & & [R$_{\odot}$] & [M$_{\odot}$] & & [\kms] & &  & & & & \p[dex] 
}
\startdata
11 & B0 I    & 27.5 & 3.05 & 35.5 & 51.6  &   5.81 & 12 & 0.10 & $<$7.3\pl & 7.8 & 6.9 & 6.7 & $-$0.9 \\   
7  & B0-1 Ia & 27.0 & 2.90 & 37.0 & 39.7  &   5.82 & 12 & 0.10 & $<$7.5\pl & 7.8 & 7.0 & 6.9 & $-$0.9 \\
9  & B0.5 Ia & 25.0 & 2.75 & 34.5 & 24.4  &   5.62 & 12 & 0.10 & 7.7 & 7.8 & 7.0 & 7.0 & $-$0.9 \\   
3  & B1 Ia   & 23.5 & 2.50 & 56.2 & 36.5  & 5.94 & 15 & 0.10 &  7.4 & 7.7 & 6.8 & 6.8 & $-$1.0 \\
22 & B1 Ia   & 22.0 & 2.60 & 33.0 & 15.6  &   5.35 & 15 & 0.15 & 7.7 & 7.8 & 6.9 & 6.9 & $-$0.9 \\
37 & B2 Iab  & 20.5 & 2.70 & 28.0 & 14.3  &   5.04 & 10 & 0.15 & 7.3 & 7.6 & 6.9 & 6.9 & $-$1.1 \\    
27 & B2.5 Ia & 19.0 & 2.40 & 37.0 & 12.5  &   5.20 & 12& 0.13 & 7.7 & 7.8 & 7.1 & 7.0 & $-$0.9  \\  
28 & B2.5 Iab & 19.0 & 2.55 & 39.0 & 19.7  &   5.25 & 10 & 0.10 & 7.7 & 7.8 & 7.1 & 7.1 &  $-$0.9 \\
\enddata
\tablecomments{$\epsilon_{\rm X} =$ log(X/H)$+$12}
\tablecomments{The ratio in the final column adopts a solar abundance of $\epsilon_{\rm O}$ = 8.66 from \citet{asp05}}
\end{deluxetable}
\end{center}

\clearpage

\begin{center}
\begin{deluxetable}{lcccccccc}
\tabletypesize{\footnotesize}
\tablewidth{0pc}
\tablecolumns{9}
\tablecaption{Abundance uncertainties as a function of stellar parameters for star \#22 \label{abunerr22}}
\tablehead{
\colhead{Element} & \colhead{Upper}     & \colhead{Lower}     & \colhead{$v_{\rm turb}$} & \multicolumn{2}{c}{$\Delta$He} & 
\colhead{$\Delta Z$} & $\Delta$log$Q$ & \colhead{$\phantom{^{(c)}} \sigma~^{(c)}$}\\
                              & \colhead{limit$^{(a)}$} & \colhead{limit$^{(b)}$} & \colhead{[$-$5~\kms]}     & \colhead{[$+$0.2~dex]} & \colhead{[$-$0.2~dex]} & 
\colhead{[$+$0.2 dex]} & \colhead{[$+$0.2 dex]} &}
\startdata
N & $-$0.10 & $-$0.01 & $-$0.08 & $-$0.06 & $-$0.03 & $-$0.06 & $-$0.06 & 0.16 \\
O & $-$0.02 & $-$0.08 & $-$0.08 & $-$0.08 & $-$0.03 & $-$0.05 & $-$0.05 & 0.16 \\
Mg & $-$0.09 & \pd0.00 & $-$0.08 & $-$0.05 & $-$0.05 & $-$0.05 & $-$0.06 & 0.16 \\
Si & $-$0.04 & $-$0.01 & $-$0.09 & $-$0.07 & \pd0.03 & $-$0.03 & $-$0.06 & 0.14 \\
\enddata
\tablecomments{\\
$^a$ T$_{\rm eff}$ $+$ 1,000~K; log$g$ $+$ 0.1 dex \\
$^b$ T$_{\rm eff}$ $-$ 1,000~K; log$g$ $-$ 0.1 dex\\
$^c$ The global abundance uncertainty ($\sigma$) is obtained as the square root of the sum of the squares of the uncertainties.
}
\end{deluxetable}
\end{center}

\begin{center}
\begin{deluxetable}{lcccccccc}
\tabletypesize{\footnotesize}
\tablewidth{0pc}
\tablecolumns{9}
\tablecaption{Abundance uncertainties as a function of stellar parameters for star \#37 \label{abunerr37}}
\tablehead{
\colhead{Element} & \colhead{Upper}     & \colhead{Lower}     & \colhead{$v_{\rm turb}$} & \multicolumn{2}{c}{$\Delta$He} & 
\colhead{$\Delta Z$} & $\Delta$log$Q$ & \colhead{$\phantom{^{(c)}} \sigma~^{(c)}$}\\
                              & \colhead{limit$^{(a)}$} & \colhead{limit$^{(b)}$} & \colhead{[$-$5~\kms]}     & \colhead{[$+$0.2~dex]} & \colhead{[$-$0.2~dex]} & 
\colhead{[$+$0.2 dex]} & \colhead{[$+$0.2 dex]} &}
\startdata
N & $-$0.05 & $-$0.08 & $-$0.09 & $-$0.08 & $-$0.02 & $-$0.05 & $-$0.05 & 0.17 \\
O & \pd0.04 & $-$0.14 & $-$0.09 & $-$0.09 & $-$0.01 & $-$0.03 & $-$0.05 & 0.20 \\
Mg & $-$0.10 & \pd0.05 & $-$0.11 & $-$0.04 & $-$0.02 & $-$0.03 & $-$0.03 & 0.16 \\
Si & \pd0.08 & $-$0.13 & $-$0.13 & $-$0.08 & \pd0.04 & \pd0.00 & $-$0.04 & 0.22 \\
\enddata
\tablecomments{\\
$^a$ T$_{\rm eff}$ $+$ 1,000~K; log$g$ $+$ 0.1 dex \\
$^b$ T$_{\rm eff}$ $-$ 1,000~K; log$g$ $-$ 0.1 dex\\
$^c$ The global abundance uncertainty ($\sigma$) is obtained as the square root of the sum of the squares of the uncertainties.
}
\end{deluxetable}
\end{center}

\clearpage

\clearpage
\begin{figure*}
\begin{center}
\caption{FORS $V$-band pre-imaging of NGC\,3109, with our spectroscopic targets encircled.  North is at the top, 
east to the left.  For reference the three saturated stars at the south-west of the frame are those just south of the `F2' 
label in Figure~1 of \citet{b93}.  The image is roughly 4\farcm0 squared.}
\label{targets1}
\end{center}
\end{figure*}

\clearpage
\begin{figure*}
\begin{center}
\caption{Target stars in NGC\,3109, moving eastwards from Figure~\ref{targets1}.}

\label{targets2}
\end{center}
\end{figure*}

\clearpage
\begin{figure*}
\begin{center}
\caption{Target stars in NGC\,3109, moving eastwards from Figure~\ref{targets2}.  To include 
\#17 the image shown here is approximately 3\farcm0$ \times $5\farcm0.}
\label{targets3}
\end{center}
\end{figure*}

\clearpage
\begin{figure*}
\begin{center}
\includegraphics[scale=0.8]{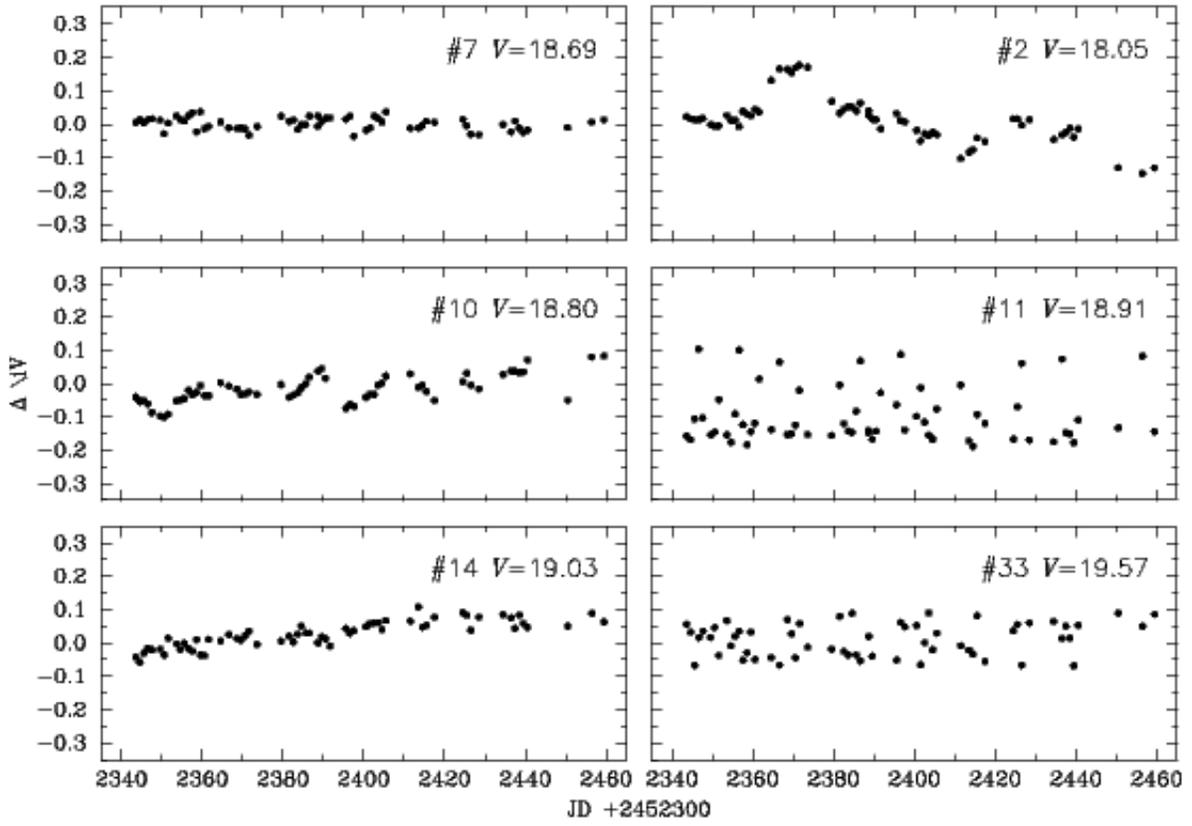}
\caption{Photometric variables in NGC\,3109.  Star \#7 is included to illustrate the scatter for
a non-variable target.}
\label{var}
\end{center}
\end{figure*}

\clearpage
\begin{figure*}
\begin{center}
\includegraphics[scale=0.8]{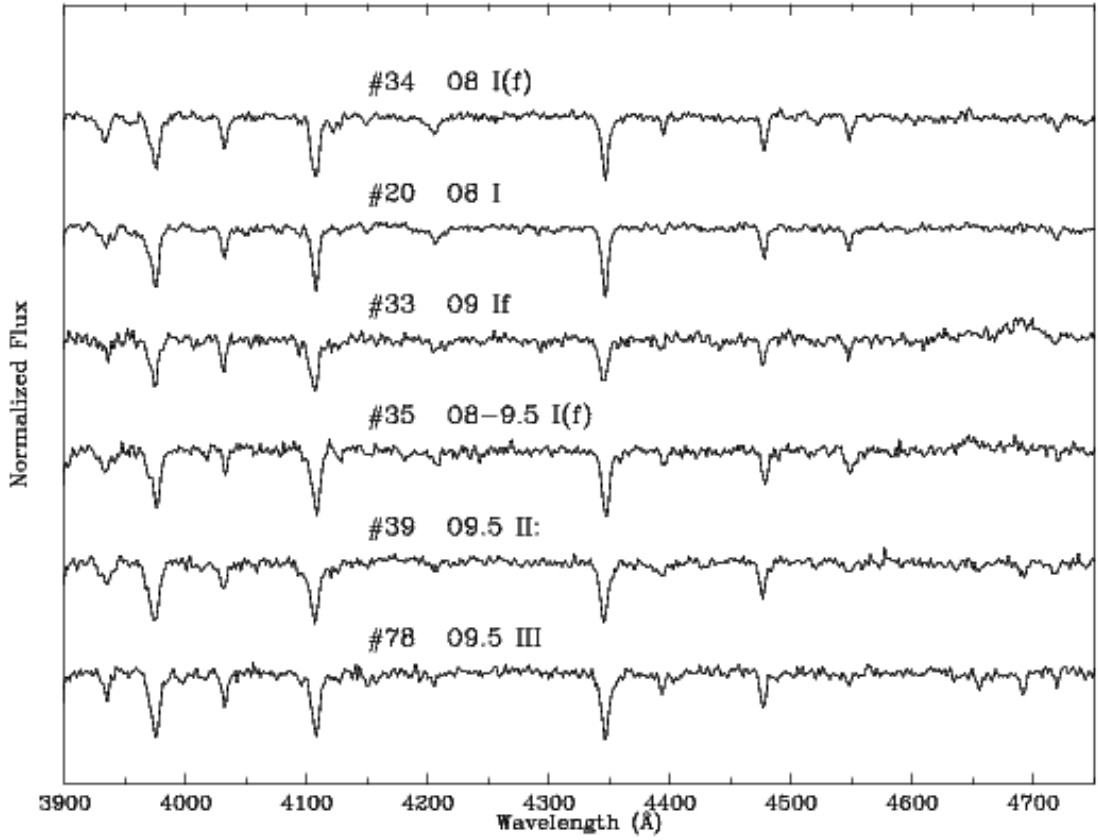}
\caption{Normalized spectra of O-type stars in NGC\,3109.}
\label{ostars}
\end{center}
\end{figure*}

\clearpage
\begin{figure*}
\begin{center}
\includegraphics[scale=0.8]{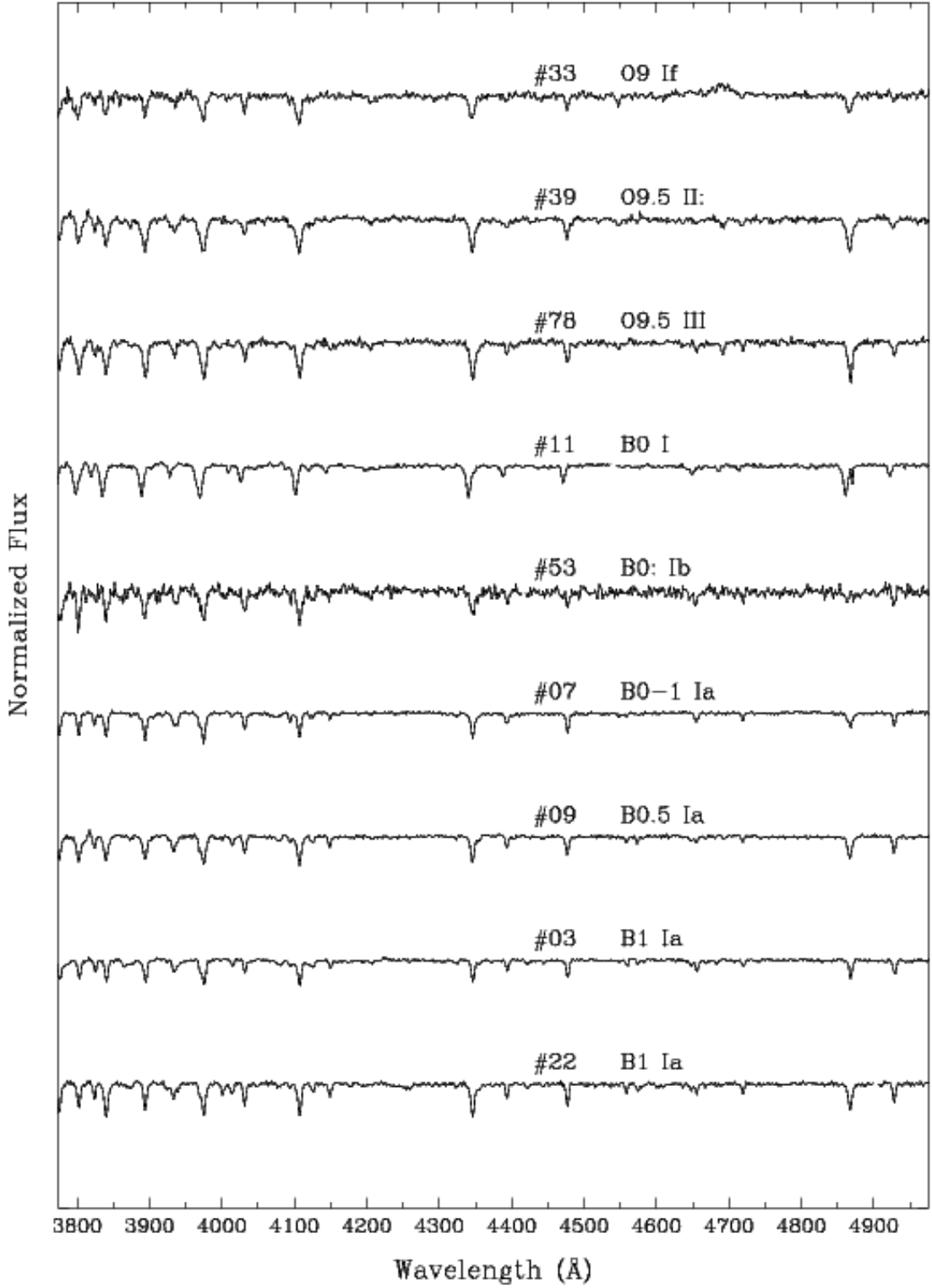}
\caption{Normalized spectra of stars classified in the range O9 to B1.}
\label{b0}
\end{center}
\end{figure*}

\clearpage
\begin{figure*}
\begin{center}
\includegraphics[scale=0.8]{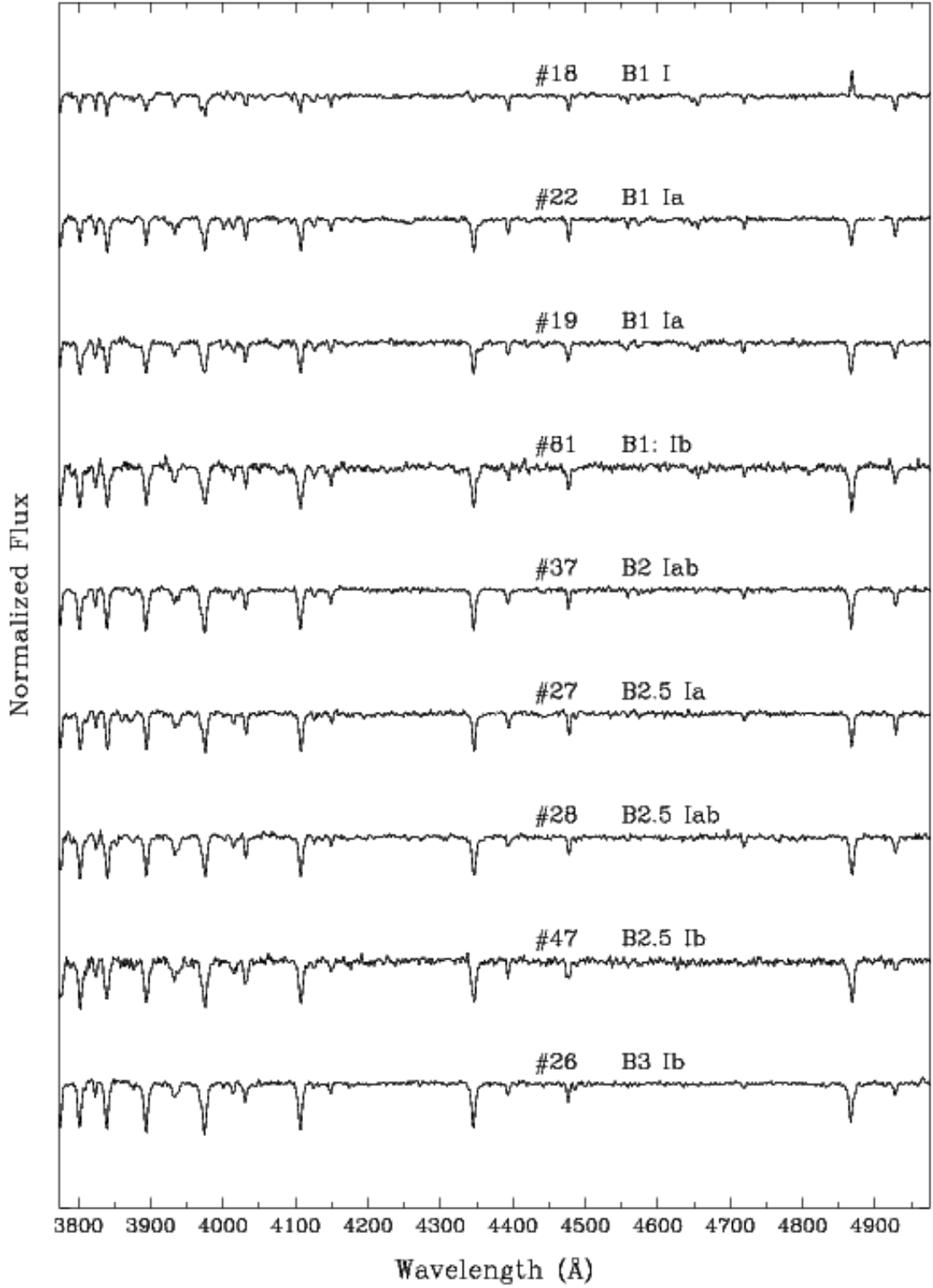}
\caption{Normalized spectra of early B-type supergiants.}
\label{bearly}
\end{center}
\end{figure*}

\clearpage
\begin{figure*}
\begin{center}
\includegraphics[scale=0.8]{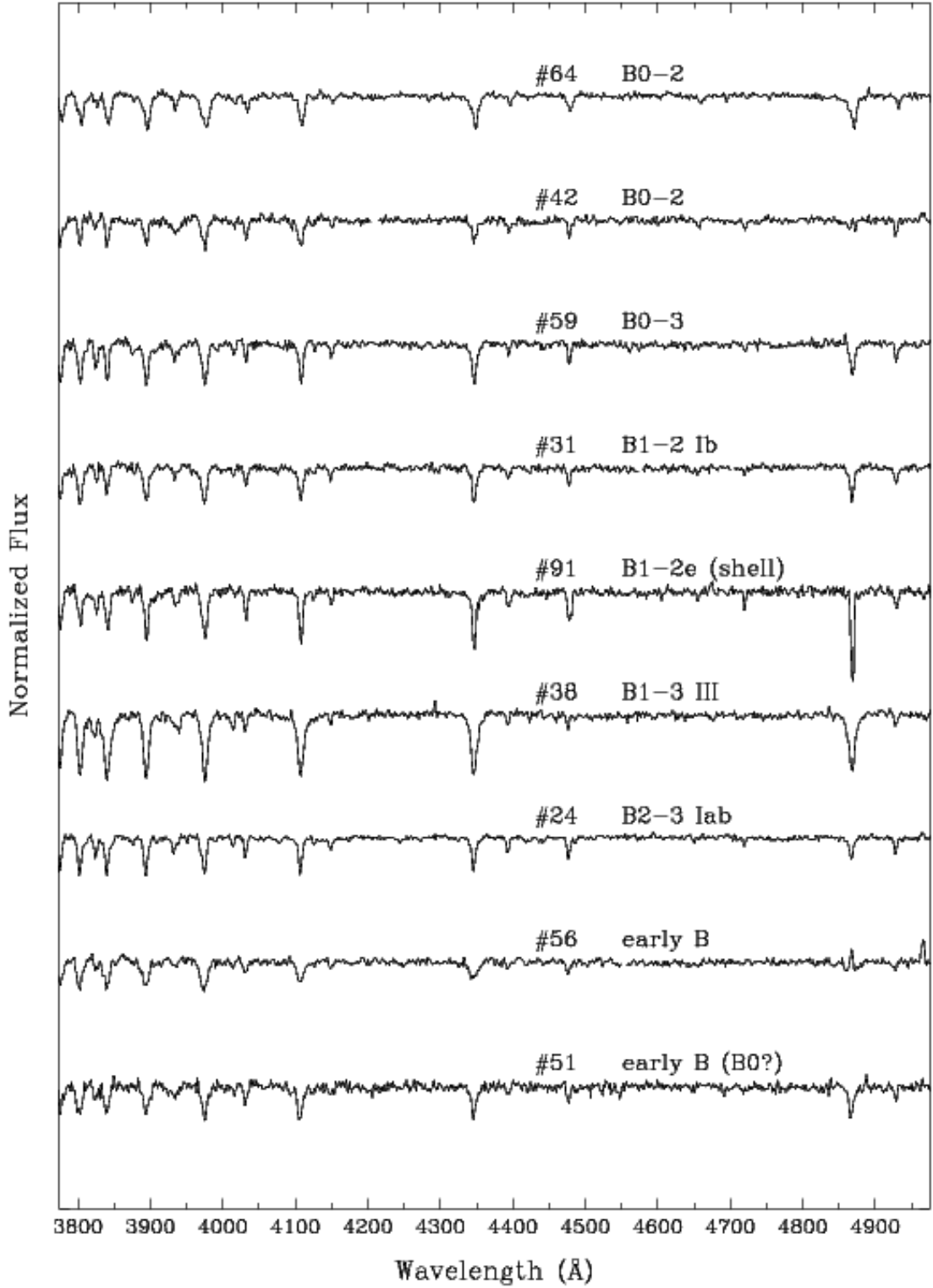}
\caption{Normalized spectra of early B-type supergiants with non-unique classifications.}
\label{bnon}
\end{center}
\end{figure*}

\clearpage
\begin{figure*}
\begin{center}
\includegraphics[scale=0.8]{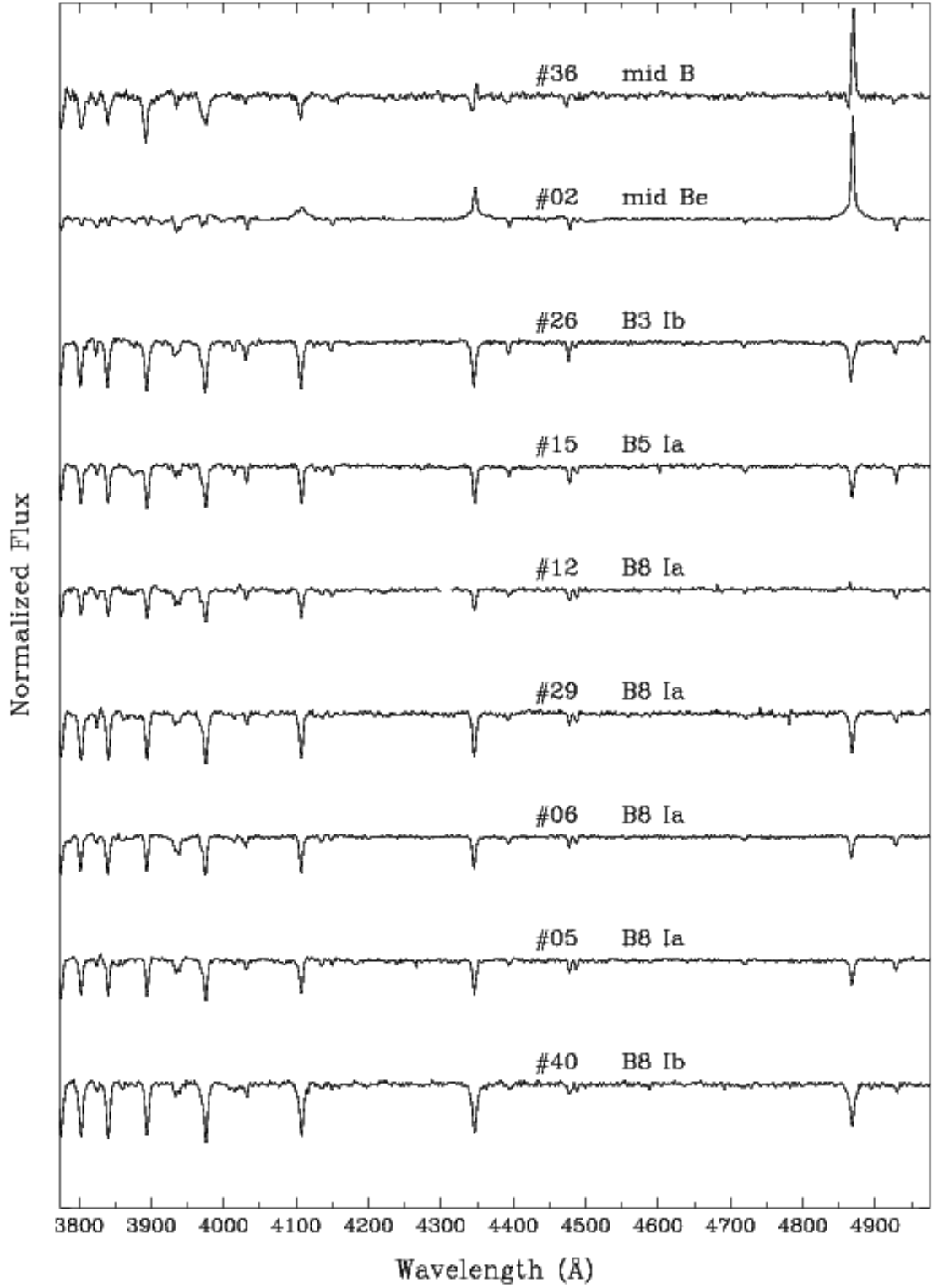}
\caption{Normalized spectra of mid B-type supergiants.}
\label{bmid}
\end{center}
\end{figure*}

\clearpage
\begin{figure*}
\begin{center}
\includegraphics[scale=0.8]{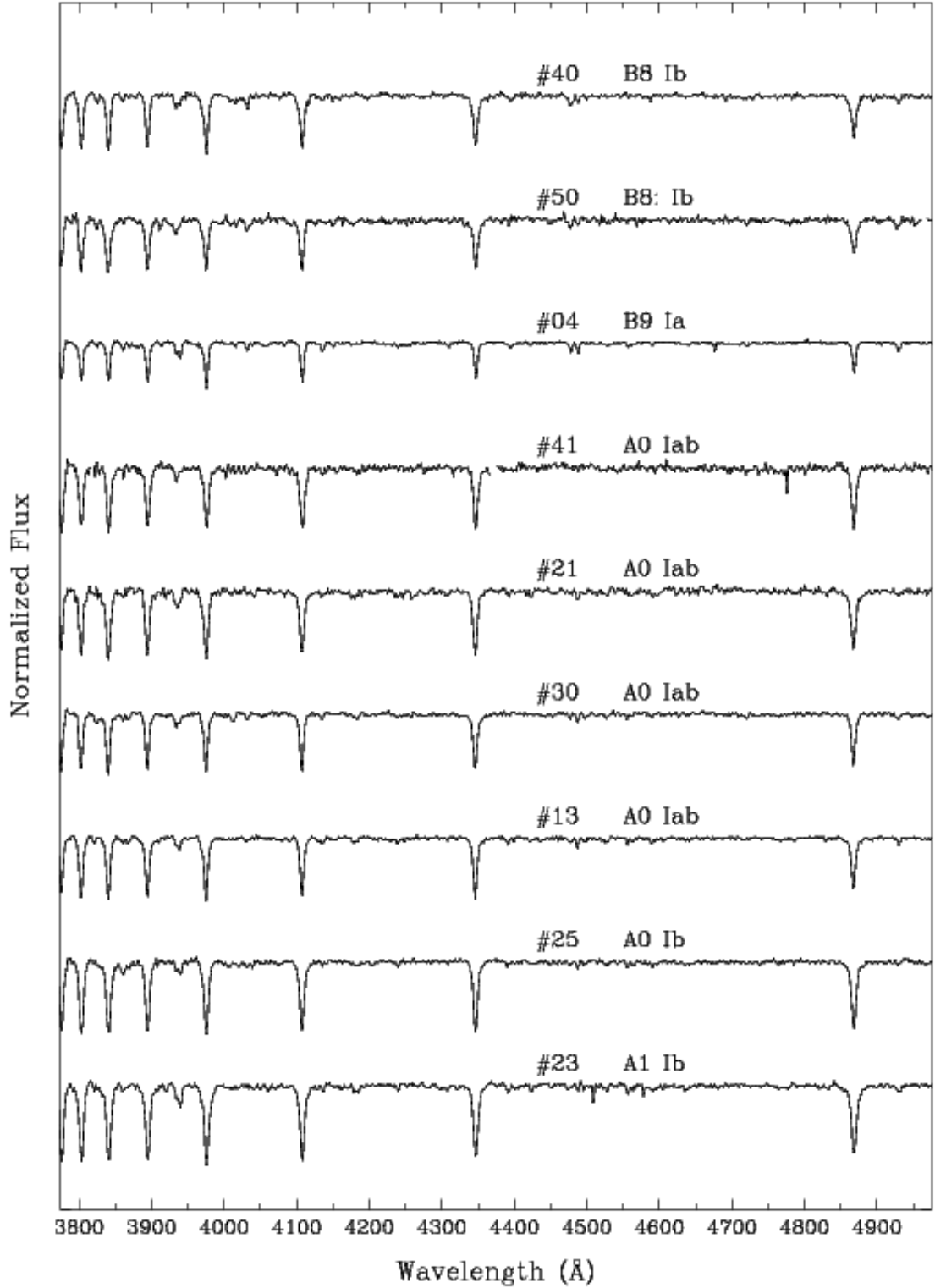}
\caption{Normalized spectra of late B- and A0-type supergiants.}
\label{blate}
\end{center}
\end{figure*}

\clearpage
\begin{figure*}
\begin{center}
\includegraphics[scale=0.8]{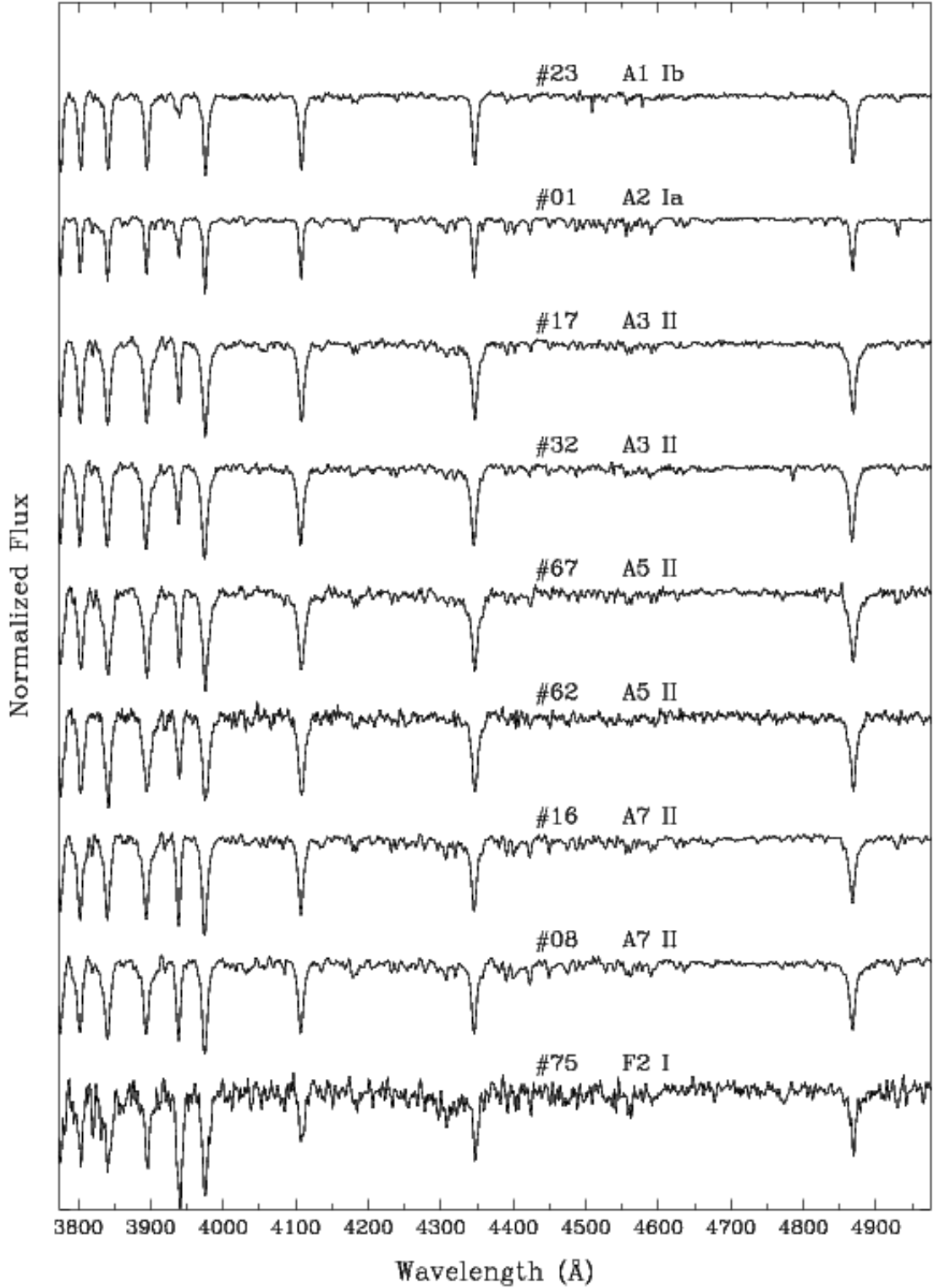}
\caption{Normalized spectra of A-type supergiants and bright giants.}
\label{astars}
\end{center}
\end{figure*}

\clearpage
\begin{figure*}
\begin{center}
\includegraphics{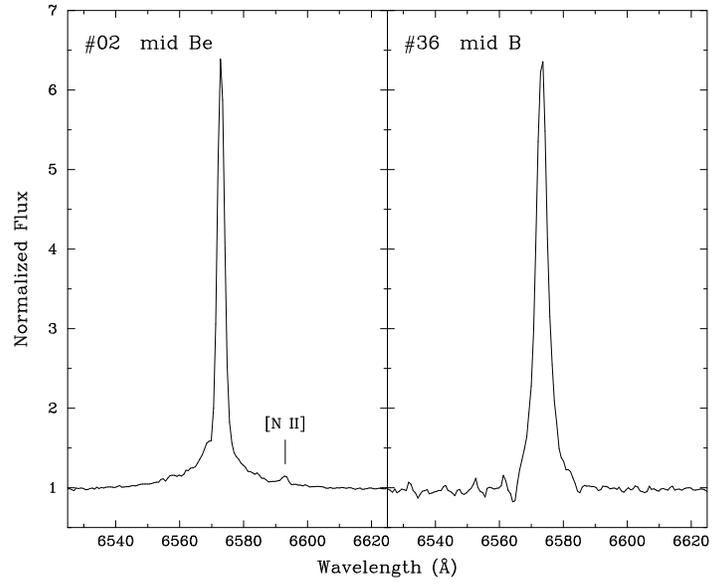}
\caption{H$\alpha$ spectra of two stars with significant emission profiles ($>$20~\AA).  The
redshifted [N~{\scriptsize II}] emission at \lam6583 is the small `bump' redward of H$\alpha$ 
in the spectrum of \#02.}
\label{red1}
\end{center}
\end{figure*}

\clearpage
\begin{figure*}
\begin{center}
\includegraphics[scale=0.9]{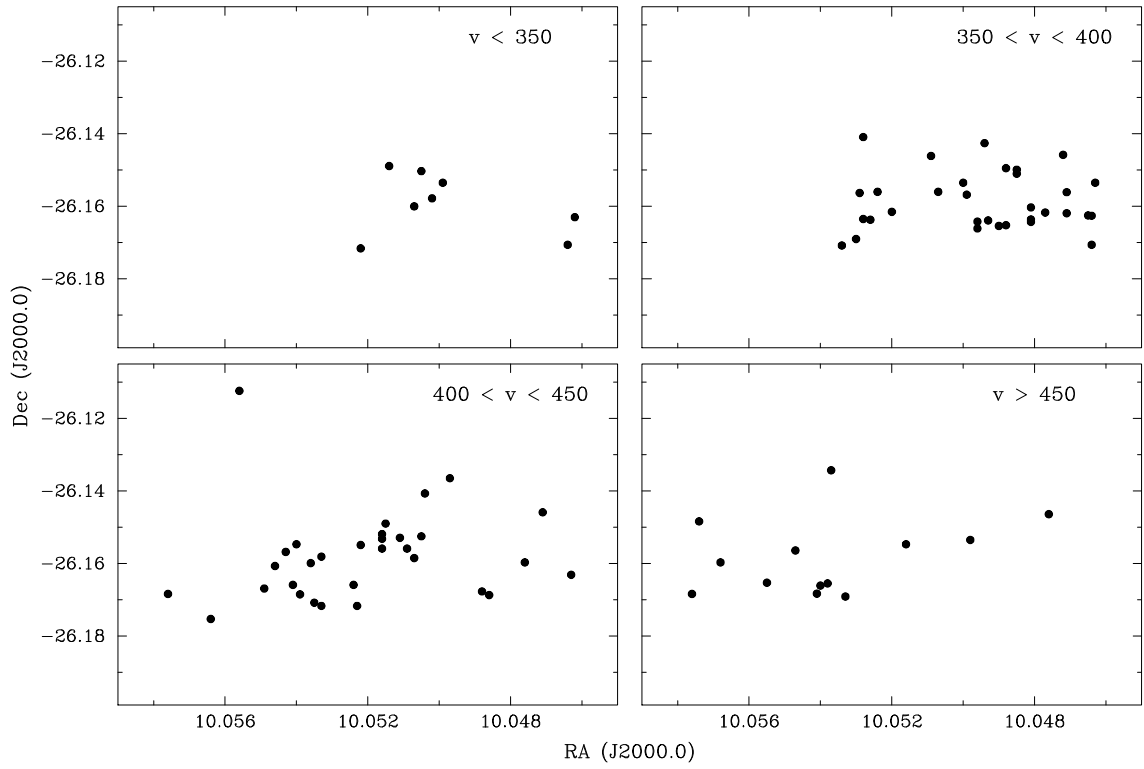}
\caption{Spatial distribution of NGC\,3109 targets in different radial velocity bins (in \kms), indicating a velocity
gradient in the kinematics of the young stellar population across the
observed region.}
\label{rvfig}
\end{center}
\end{figure*}

\clearpage
\begin{figure*}
\begin{center}
\includegraphics[scale=0.9]{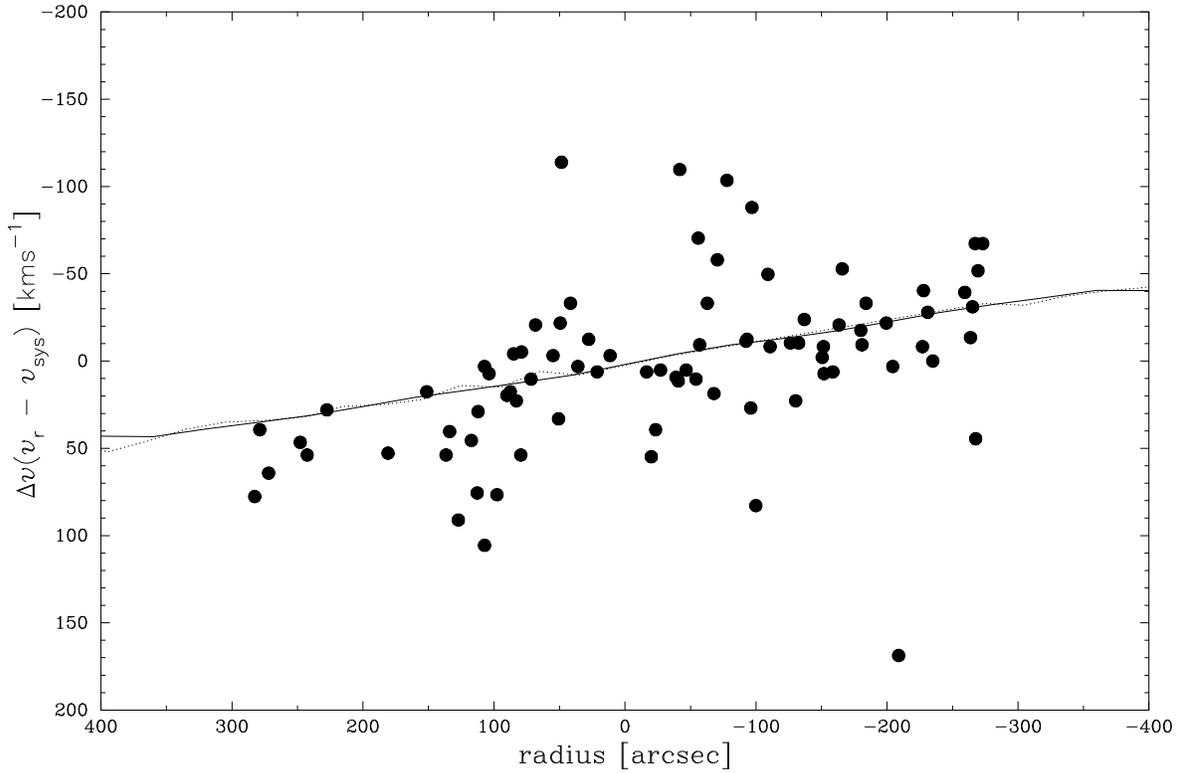}
\caption{Stellar radial velocities, as a function of radius along the major axis of NGC\,3109, 
corrected to the systemic velocity a 402~\kms.  The median 1-$\sigma$
(internal) uncertainty is 19~\kms.  Also shown are rotation curves
from H~\1 \citep[][solid line]{jc90} and H$\alpha$ \citep[][dotted line]{bac01}.}
\label{rvc}
\end{center}
\end{figure*}

\clearpage
\begin{figure*}
\begin{center}
\includegraphics[scale=0.7]{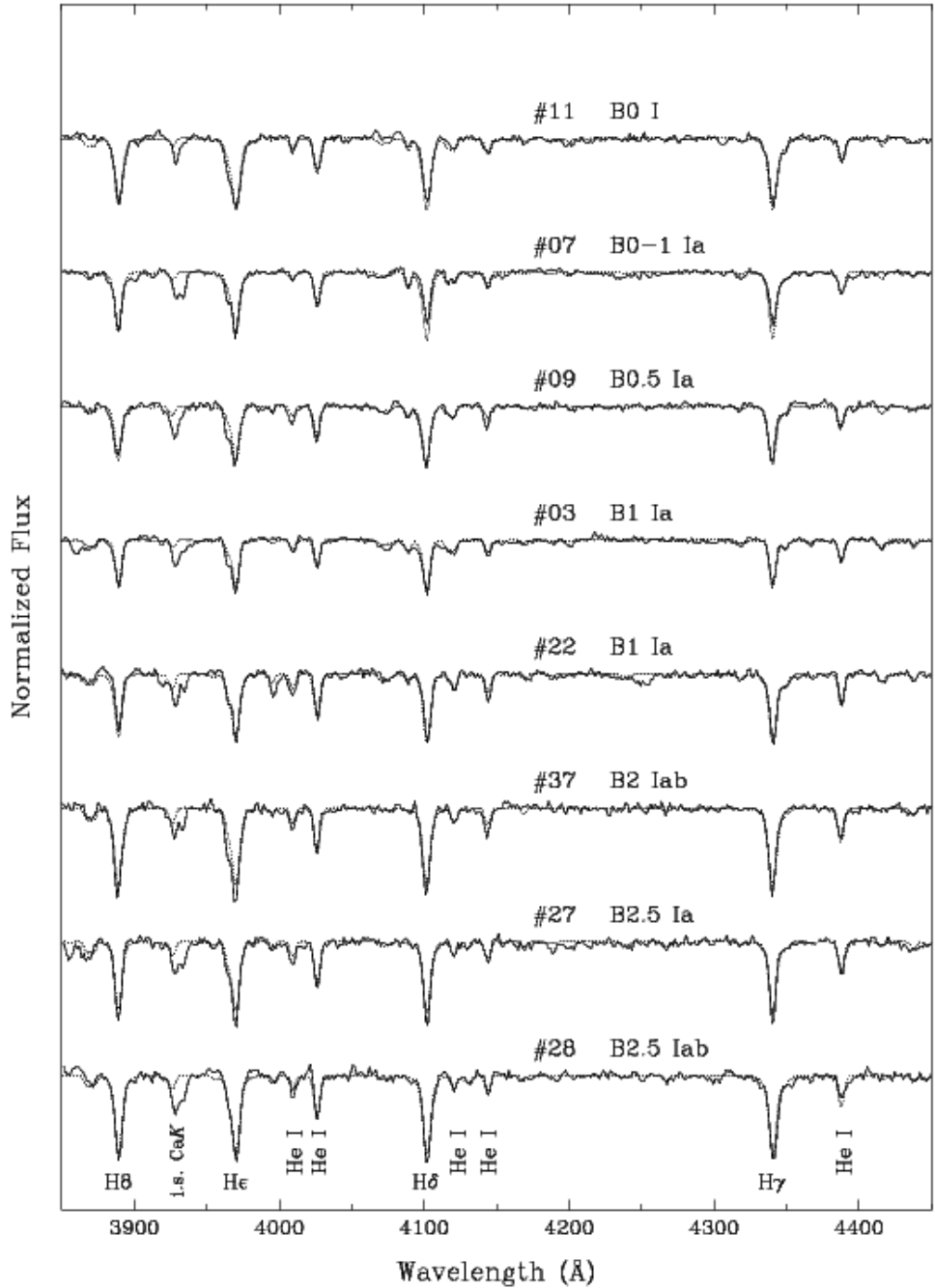}
\caption{FORS spectra (\lam\lam3850-4450 \AA) of B-type supergiants in NGC\,3109 (solid 
lines) compared with the adopted {\sc fastwind} models (dotted lines).  In addition to the Balmer
lines, the He~{\footnotesize I} \lam\lam4009, 4026, 4121, 4143, 4388 lines are marked.  Most of
the absorption $\sim$\lam3933~\AA\/ arises from the interstellar Ca~$K$ line.}
\label{fits1}
\end{center}
\end{figure*}

\clearpage
\begin{figure*}
\begin{center}
\includegraphics[scale=0.7]{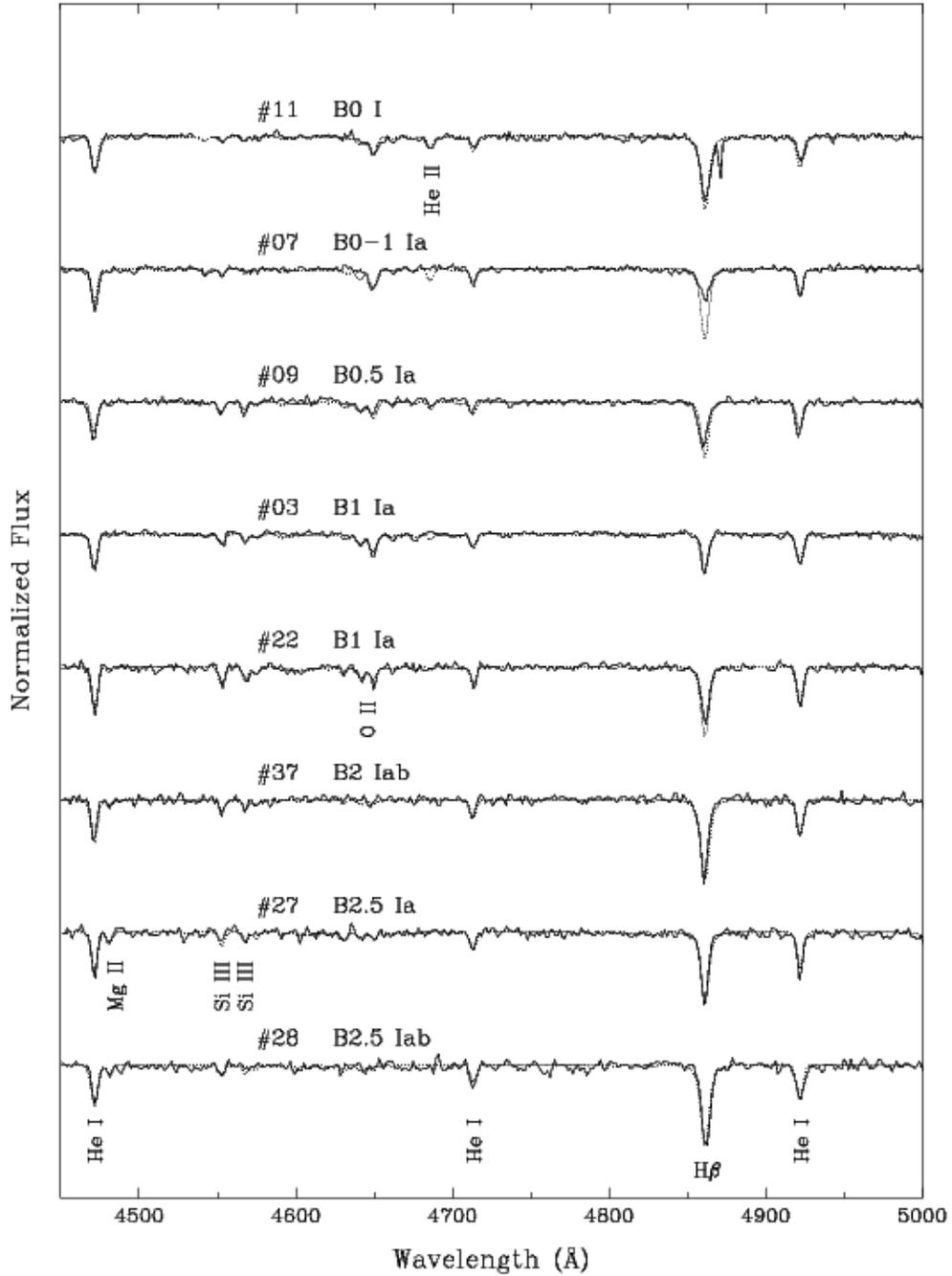}
\caption{FORS spectra (\lam\lam4450-5050 \AA)  of B-type supergiants in NGC\,3109 (solid 
lines) compared with the adopted {\sc fastwind} models (dotted lines).  The identified 
absorption lines in the spectrum of star \#28 are He~{\footnotesize I} \lam\lam4471, 4713, 4922, 
and H$\beta$; star \#27: Mg~{\footnotesize II} \lam4481, Si~{\footnotesize III} \lam\lam4553-4567; 
star \#22: O~{\footnotesize II} \lam\lam4640-4650; star \#11: He~{\footnotesize II} \lam4686.
}
\label{fits2}
\end{center}
\end{figure*}

\clearpage
\begin{figure*}
\begin{center}
\includegraphics[scale=0.75]{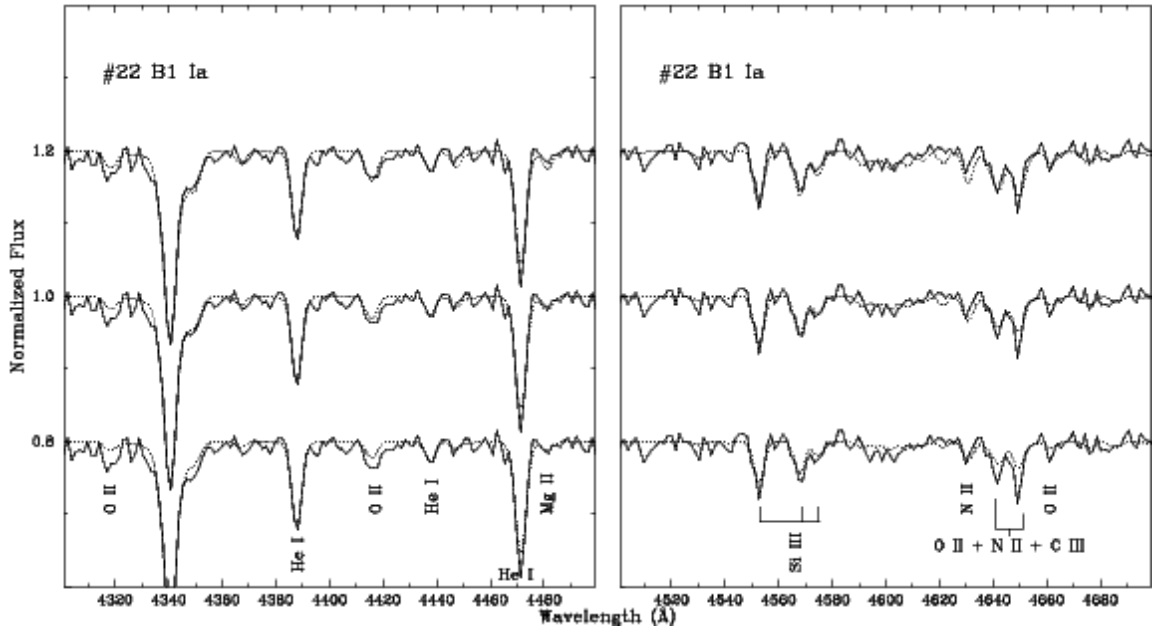}
\caption{Comparison of FORS spectrum of star \#22 (solid lines) with {\sc fastwind} models in which the 
metallic abundances are $\pm$0.2~dex compared to the adopted values (upper and lower spectra respectively).
The identified lines are O~{\footnotesize II} \lam\lam4317-19, 4415-17, 4661; 
He~{\footnotesize I} \lam\lam4388, 4438, 4471; Mg~{\footnotesize II} \lam4481; Si~{\footnotesize} \lam\lam4552-68-75; 
N~{\footnotesize II} \lam4631; and the O~{\footnotesize II}~$+$N~{\footnotesize II}~$+$C~{\footnotesize III} blend at
\lam4650. \label{err_22} }
\end{center}
\end{figure*}

\begin{figure*}
\begin{center}
\includegraphics[scale=0.75]{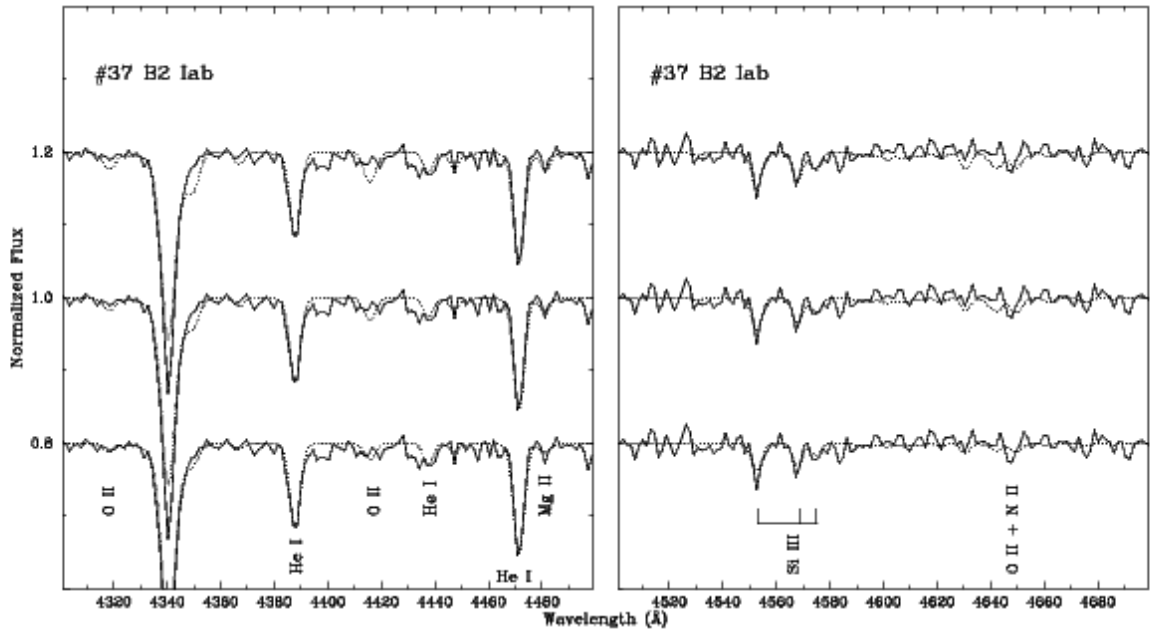}
\caption{Comparison of FORS spectrum of star \#37 (solid lines) with {\sc fastwind} models in which the 
metallic abundances are $\pm$0.2~dex compared to the adopted values (upper and lower spectra respectively). \label{err_37}}
\end{center}
\end{figure*}

\end{document}